\documentclass[preprint,3p,number]{elsarticle}
\pdfoutput=1
\usepackage[T1]{fontenc}
\usepackage[utf8]{inputenc} 
\usepackage{amsmath}
\usepackage{amsmath}
\usepackage{amsfonts}
\usepackage{amssymb}
\usepackage{latexsym}
\usepackage{graphics}
\usepackage{graphicx}
\usepackage{hyperref}
\usepackage{epigraph} 
\usepackage{framed} 
\usepackage{epsf}
\usepackage{aniszewski,color}
\usepackage[hang,footnotesize,bf]{caption} 
\title{Volume of Fluid (VOF) type advection methods in two-phase flow: a comparative study.}

\author[coria]{W. Aniszewski\corref{cor1}}  
\ead{aniszewski@coria.fr}

\cortext[cor1]{Corresponding Author. email: aniszewski@coria.fr; tel. +33 (0)2 32 95 36 73,  CNRS-UMR6614 CORIA, Universite de Rouen, Site Universitaire du Madrillet - BP 12, 76801 Saint Etienne du Rouvray cedex}

\author[coria]{T. M\'{e}nard}
\author[imc]{M. Marek}

\address[coria]{CNRS UMR 6614 - CORIA Rouen, Site Universitaire du Madrillet, Saint Etienne du Rouvray, France}
\address[imc]{Institute of Thermal Machinery, Czestochowa University of Technology, Czestochowa, Poland}

\journal{Computers \& Fluids}

\newcommand{\specialcell}[2][c]{%
  \begin{tabular}[#1]{@{}c@{}}#2\end{tabular}}

\begin{document}
\bibliographystyle{plain}

\begin{abstract}
  In this paper, four distinct approaches to Volume of Fluid (VOF) computational method are compared. Two of the methods are the 'simplified' VOF formulations, in that they do not require geometrical interface reconstruction. The assessment is made possible by implementing all four approaches into the same code as a switchable options. This allows to rule out possible influence of other parts of numerical scheme, be it  the discretisation of Navier-Stokes equations or chosen approximation of curvature, so that we are left with conclusive arguments because only one factor differs the compared methods. The comparison is done in the framework of CLSVOF (Coupled Level Set Volume of Fluid), so that all four methods are coupled with Level Set interface, which is used to compute pressure jump via the GFM (Ghost-Fluid Method). Results presented include static advections, full N-S solutions in laminar and turbulent flows. The paper is aimed at research groups who are implementing VOF methods in their computations or intend to do it, and might consider a simplified approach as a preliminary measure, since the methods presented differ greatly in complication level, or ease of implementation expressed e.g. in number of code lines.
\end{abstract}


\maketitle

\section{Introduction}
In this paper, a comprehensive, comparative study of four approaches to Volume of Fluid (VOF) methods is presented.  Using also Level Set (LS) method \cite{osher2000}, the comparison is performed in the framework of CLSVOF (Coupled Level Set Volume of Fluid), i.e. four VOF methods are coupled with LS and tested for the simulations of physical flows. Additional set of passive advections is provided performed both in CLSVOF manner as well as ``pure'' VOF approach. Tested approaches are THINC/SW (Tangent of Hyperbola Interface Capturing with Slope Weighting) \cite{xiao2011,xiao} (here, for shortening notation,  referred to as '\textbf{M2}'), the WLIC (Weighted Linear Interface Calculation) \cite{yokoi,svof} (referred to as '\textbf{M3}'), the PLIC (Piece-wise Linear Interface Calculation) \cite{tsz, aniszewskiJCP} (referred to as '\textbf{M4}') and a CLSVOF method using interface reconstruction techniques \cite{menard} (which we designate '\textbf{M1}').

While the latter two approaches involve full geometrical reconstruction of the interface (and, as a consequence, are relatively complicated from programmer's point of view), the former two have been proposed as simplified versions of ''full'' VOF method. Assessment of the results suggests reasonable dependence between sophistication (expressed e.g. in number of FORTRAN code lines)  and accurateness of the method, when considering  parameters such as the obtained mass conservation and interface smoothness. The analysis also includes a rather non-standard view on the methods from the programmer's man-hour point of view. The authors claim that presented comparisons stand out in that, to their knowledge, this is the first study in which only one single Navier-Stokes solver code  was utilized, having all the methods implemented available as switchable run-time options.

In this paper, we give a description of the VOF method in general; followed by very brief characterizations of its four tested variants. The coupling between VOF and Level Set \cite{osher88, osher2000} methods as well the numerical characteristics of utilized Navier-Stokes solver are only described in shortened form, since they would go far beyond the scope of this paper; instead, the reader is pointed to accurate literature sources. The numerical experiments section follows along with the assessment of the tested methods.

Several years ago, a similar comparative work has been  published by Gerlach et al. \cite{gerlach2005}. However, the authors of that paper have performed tests of advanced methods (involving LVIRA, PROST and CLSVOF schemes for the interface reconstruction) which were not ``simplified'' in the sense \textbf{M2} and \textbf{M3} methods are. Also, \cite{gerlach2005} contained solely two-dimensional tests of flows (rising bubble simulations). We enlarge the scope by including three-dimensional simulations, and focus on the  ''simplified'' methods that have appeared since (as is the case of \textbf{M3}) or have been recently re-visited by their authors (\textbf{M2}, see \cite{xiao2011}). Thus, these methods are not obsolete (see e.g. recent publication \cite{yokoi2013} concerning \textbf{M3}) and testing their capabilities is justified.
\section{Description of tested VOF methods}
\subsection{General VOF description}

In VOF method \cite{hn,tsz} the fraction (color) function $C$ is used which is an integral of phase's characteristic function, in $\mathbf{R^2}$:
\begin{equation}\label{sv0}
  C_{ij}=\frac{1}{\Delta x \Delta y}\int\limits_0^{\Delta x}\int\limits_0^{\Delta y}\chi(x,y)dxdy.
\end{equation}

Since $\chi$ is passively advected, its substantial derivative is zero

\begin{equation}\label{sv01}
  \frac{D\chi}{D t}=0.
\end{equation}

We assume that the velocity field is incompressible,

\begin{equation}\label{sv1}
  \nabla\cdot\ub=0
\end{equation}

which enables us to put

\begin{equation}\label{sv2}
  \frac{\partial\chi}{\partial t}+\frac{\partial \chi u}{\partial x}+\frac{\partial \chi v}{\partial y}=0.
\end{equation}

To discretize the equations, one uses the definition (\ref{sv0}), and integration in the control volume using Ostrogradski-Gauss theorem \cite{tsz}. Products $\chi u,$ $\chi v$ represent fluxes of $\chi$ through the control volume boundaries. The discretized form of this equation has the form:

\begin{equation}\label{sv3}
  C_{ij}^{n+1}=C^n_{ij}+\frac{\Delta t}{\Delta x}\left((F_x)^n_{i-1/2,j} -(F_x)^n_{i+1/2,j}\right)+\frac{\Delta t}{\Delta y} \left( (F_y)^n_{i,j-1/2}-(F_y)^n_{i,j+1/2}\right),
\end{equation}

in which the term $(F_x)_{i+1/2,j}$ denotes the flux through the right-hand wall ($x$ direction) of grid cell $(i,j)$ with dimensions $\Delta x, \Delta y.$ The staggered grid is used, so this term is calculated using the $u$ value taken at the same point, in a manner described below. The (\ref{sv3}) formula applies to a \textit{split} approach towards VOF advection \cite{szdirect}, i.e. calculate fluxes separately along axes and perform three separate advections.

The necessity of using flux expressions $F_x, F_y$ (and $F_z$ in $\mathbf{R}^3$) stems from the discontinuous character of the fraction function $C,$  due to which the (\ref{sv01}) equation cannot be solved directly using $C$ concept\footnote{As opposed to Level Set \cite{osher2000} methods, where the distance function $\phi$ is continuous and its advection equation is solved directly}. The calculation of the VOF flux terms is done geometrically in full-fledged VOF implementations such as PLIC \cite{szdirect}. Two simplified VOF approaches tested in this article introduce simplified expressions to solve (\ref{sv3}). To facilitate the understanding of the simplified THINC and WLIC methods, we follow with a description of the PLIC approach.

\subsection{PLIC method}

The PLIC (Piece-wise linear Interface Calculation) \cite{youngs92, szdirect} approach to VOF is based upon the following concepts.  The fluxes $F^n$ of the fraction function are calculated geometrically as volumes of fluid which get exchanged between the cells. It is assumed, that the interface is a line/plane, and it can be described by the equation

\begin{equation}\label{pl0}
  \nb\xb=\alpha
\end{equation}

in Eulerian space. To simplify the formulae we further assume that the normal vector is unitary, that is $|\nb |=1,$ and that uniform discretisation $\Delta x=\Delta y=h$ is used. All these assumptions can be easily incorporated into computational codes. In PLIC, as described e.g. in \cite{pilliod}, the $\nb$ can be obtained using a differential scheme (see e.g. Youngs \cite{youngs1991})  It has been shown, that this approximation is at most 1st order \cite{pilliod}. Thus, errors - especially in non-resolved areas  - are expected with this type of scheme, as far as the shape of the reconstructed interface is considered. The same goes for the curvature $\kappa$ calculated when using $\nb,$  if such an approach is chosen to calculate the surface tension \cite{popinet1}. Second-order accurate alternatives exist \cite{pilliod, miller}, and such an approach \cite{menard} is used in this paper in the \textbf{M1} and \textbf{M4}. To make the comparison reliable, all other methods were implemented using the same normal vector calculating routines (if the normal vector is needed).

\begin{figure}[ht!]
  \centering
  \includegraphics[scale=1.]{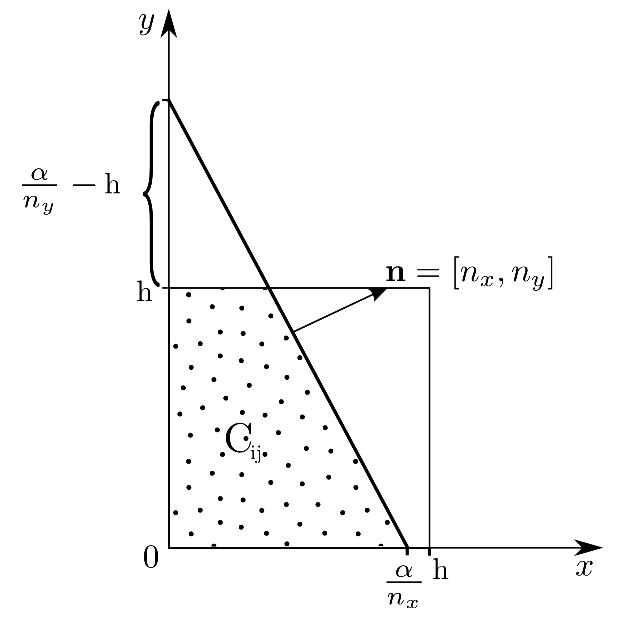}
  \caption{Example of an interface grid cell in $\mathbf{R}^2$ with a PLIC interface reconstruction.}\label{plicf1}
\end{figure}

To fully describe the interface equation (\ref{pl0}), the free term $\alpha$ is needed. Its analytic calculation is possible on a geometrical basis \cite{aniszewski} by enforcing the mass conservation in the cell. In $\mathbf{R}^2,$ the area $V$ enclosed by the interface when $n_x,n_y>0$ is equal to

\begin{equation}\label{p3}
  V=\frac{\alpha}{2n_x n_y},
\end{equation}

provided that the points $\alpha/n_x$ and $\alpha/n_y$ lie inside the cell, so \[ \frac{\alpha}{n_{x,y}}\le h. \] If this condition is not fulfilled (so for example $\alpha/n_x>h$) appropriate triangular areas (see Fig. \ref{plicf1})  must be subtracted from right-hand side of (\ref{p3}) to get $V.$ Once the relation $V(\alpha/h)$ is known, and knowing also that \begin{equation}\label{p31}V_{ij}(\alpha)=C_{i,j}\end{equation},  the $\alpha$ can be found algebraically. Although not trivial, the relations have been published e.g. by Scardovelli and Zaleski \cite{sz2000}.

The PLIC variant presented in this article (method \textbf{M4}) uses the analytic calculation of $\alpha,$ also in 3D. Numerical approximations of $\alpha$ using error minimization methods can also be  applied to (\ref{p31}) and so is the case of the \textbf{M1} method\footnote{In methods \textbf{M2} and \textbf{M3} usage of  $\alpha$ is not necessary.}.

\begin{figure}[ht!]
  \centering
  \includegraphics[scale=1.]{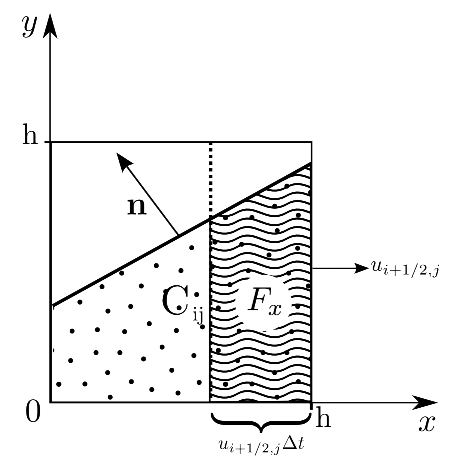}
  \caption{An example of a flux calculation problem in two dimensions.}\label{plicf2}
\end{figure}

Calculation of the fluxes $F_{x,y,z}$ is done by finding geometrically (analytically) the areas/volumes of the intersections between the interface described by (\ref{pl0}) and cuboid $dz\cdot dy\cdot u_{i+1/2,j}dt$ (for $x$ direction in $\textbf{R}^3$) \cite{aniszewski}. In Figure \ref{plicf2}, a simple two-dimensional illustration of the problem is presented, with marked $(F_x)_{i+1/2,j}$ area. Figure \ref{wlicf1} presents a simplified case with only one non-zero normal component.  It is important to note that finding these intersections areas/volumes is nontrivial matter, because a lot of interface positions need to be considered. Arguably, this stage of the VOF implementation is the hardest to implement numerically. The code used to solve (\ref{p31}) may be re-adapted for this \cite{aniszewskiJCP}, with appropriate transformations used to reduce the intersection finding problem to solving an equation equivalent to (\ref{p3}).

Even considering the possible programming amelioration,  the implementation of flux computations may require many lines of programming code, in our case the number varies from few hundred lines (\textbf{M4}) to around 1600  in \textbf{M1} (when counting interface reconstruction + flux computations).

\subsection{WLIC method}

The WLIC (Weighted Linear Interface Calculation) method has been published by K. Yokoi \cite{yokoi}, also independently by M. Marek et al. \cite{svof} as ``SVOF'' (Simplified Volume of Fluid). Published applications include both passive advections as modelling droplet splashes using CLSVOF approach \cite{yokoi2013}. As we show below, this method can be seen as  a whole family of methods, among which the two published implementation can be placed. WLIC is a method distinguishing itself from PLIC because there is no necessity of full image reconstruction. Namely, in (\ref{pl0}) free term $\alpha$ is not needed. Consequently,  calculation  of flux (\ref{sv3}) is done in a simplified way compared to PLIC, requiring no complex geometrical definition of flux areas/volumes.

\begin{figure}[ht!]
  \centering
  \includegraphics[scale=1.]{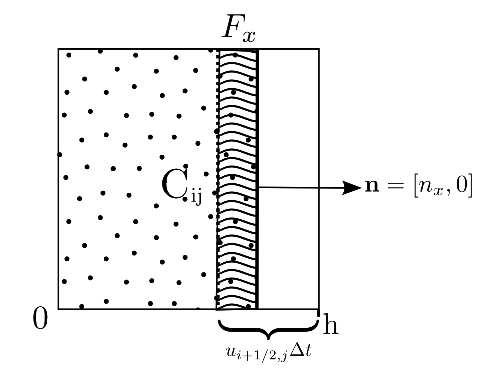}
  \caption{An example of the flux computation problem for an interface with zero $n_y$ component, as used in the WLIC method. }\label{wlicf1}
\end{figure}

Calculation of the fluxes $F_{x,y,z}$ is trivial in case of only single component of the normal being non-zero (see Figure \ref{wlicf1}). Searching for $F_x$ in $\mathbf{R^2}$, for $n_x>0$ and $n_y=0$ we have:

\begin{equation}\label{svof1}
  F_x=\max\left(0,\frac{u_{i+1/2,j}\Delta t}{h}-(1-C_{ij})\right),
\end{equation}

while for $n_y>0$ and $n_x=0$ we get

\begin{equation}\label{svof2}
  F_x=\frac{C_{ij}u_{i+1/2,j}\Delta t}{h}
\end{equation}

assuming positive right wall velocity $u$ value\footnote{When this velocity is negative, transformations are used to reduce it to above cases - the same is done when components of $\nb$ are negative.}. These expressions date back to the donor-acceptor schemes SLIC/SOLA-VOF \cite{hn}. The founding idea of WLIC is that in general $(n_x \ne 0 \wedge n_y \ne 0)$ case, interpolation can be used to find the fluxes $F_x$  from the (\ref{svof1}) and (\ref{svof2}) type formulae. Intuitively, the oblique position of the normal vector is treated as  intermediate between vertical  and horizontal positions.  Thus, as a base of interpolation, quantities derived from $\nb$ components are used.

To put this in a strict context, we define  weights $\omega_i$ to interpolate between the ``trivial cases''. Keeping assumptions from (\ref{svof1}) and (\ref{svof2}), we have:

\begin{equation}\label{svof21}
  F_x=\begin{cases}
  \omega_i \cdot\min(u_{i+1/2,j}\Delta t,C_{ij}h) & \Leftrightarrow n_x>0 \\
  \omega_i \cdot\max(0,u_{i+1/2,j}\Delta t-(1-C_{ij})h) & \Leftrightarrow n_x<0.
  \end{cases}
\end{equation}

Naturally, for $u_{i+1/2,j}=0$ there is locally no advection, and for $n_x=0$ we get  a trivial case analogous to (\ref{svof1}).

The choice of weighting function $\omega_i$ used in (\ref{svof21}) is arbitrary: three example possibilities are listed in Table \ref{svof22}. Symbol $\nb$  there corresponds to local normal vector $\nb_{ij}.$  First two variants are used in \cite{yokoi} and \cite{svof} respectively, while the third is proposed by these authors. We have tested above variants of $\omega_i$ weight functions within the same framework while preparing \textbf{M3} implementation; the results and the choice of variant will be explained below. Using a chosen $\omega_i$ from Table (\ref{svof22}) in (\ref{svof21}) we obtain the $F_x$ flux.

\begin{table}
  \begin{center}
    \caption{Weighting function variants for the WLIC method.}
    \begin{tabular}{c|c|c|c|}

      \cline{2-4}
      & I  & II  &  III  \\
      \cline{1-4}
      \multicolumn{1}{ |c| }{ }  &               &                &                 \\
      \multicolumn{1}{ |c| }{$\omega_i=$} & $\frac{|n_x|}{|\nb|}$ & $1-\frac{2}{\pi}\arccos\left(\frac{|n_x|}{|\nb|}\right)$  & $\tanh\left(\frac{|n_x|}{|\nb|}\right)$ \\
      \multicolumn{1}{ |c| }{ } &  & & \\
      \cline{1-4}
    \end{tabular}\label{svof22}
  \end{center}
\end{table}

When extending the WLIC method to three-dimensional simulation, variant I requires no change at all, and $n_x$ is used. However, variants II and III in three dimensions use:

\begin{equation}\label{svof23}
  \frac{n_{xy}}{|\nb|}
\end{equation}
as argument of $\arccos$ and $tanh$ respectively. Numerator of (\ref{svof23}) is to be understood as a projection of $\nb$ onto a plane parallel to advection direction (local $xy$ plane). As such, variant I is simplest to implement.

As we have demonstrated, in WLIC, only $\nb$ is needed to describe the interface, so full (\ref{pl0}) equation is not required, eliminating the need of knowing  $\alpha.$ This greatly relieves the programmer by removing almost all the geometrical considerations from the code (i.e. finding $\alpha$ and $F_{x,y,z}$  fluxes as geometrical intersections). After $\nb$ has been found, flux expressions (\ref{svof21}) enable the update of $C^n$ to $C^{n+1}$ without any additional steps; the WLIC flux calculations itself can be performed in as little as $70$ lines of FORTRAN code.

\subsection{THINC method}\label{thincsection}

Note that the dependence

\begin{equation}\label{fceq}
  F_x=F_x\left(\frac{u_{i+1/2}\Delta t}{h}\right)
\end{equation}
can be plotted for trivial cases such as (\ref{svof1}) piece-wise-linearly. For non-trivial cases, the dependence becomes a quadratic curve \cite{sz2000,aniszewski}. The WLIC method approximates this curve by a sum of two lines. In contrast, the THINC (and THINC-SW) methods base on approximating the same curve with a fitted $\tanh$ functions (see section \ref{fluxcurvsect} for an example).

The THINC (Tangent of Hyperbola Interface Capturing \cite{xiao}), described first by F. Xiao et al. \cite{xiao},  in its latest incarnation has been presented in \cite{xiao2011} as THINC/SW (THINC with Slope Weighting). It is a heavily simplified VOF method which, remarkably, contains almost no geometrical interface reconstruction (\cite{xiao2011} introduces the necessity of calculating the interface normal). The authors present method formulation for the general, non-solenoidal case, so the equation

\begin{equation}\label{th1}
  \frac{\partial C}{\partial t}+\nabla\cdot(\ub C)-C\nabla \ub=0
\end{equation}
is used to represent the fraction function advection. The method is presented as strictly one-dimensional, with all the steps analogous in $y$ and $z$ directions to what is presented below. The fluxes $F_{x,y,z}$ found in (\ref{sv3}), are calculated by analytic integration of the eponymous $tanh$ function, namely

\begin{equation}\label{th2}
  \Phi_i(x)=\frac{1}{2}\left\lb 1+ \gamma \tanh \left( \beta \left(\frac{x-x_{i-1/2}}{\Delta x_i} -x_m\right)\right)\right\rb.
\end{equation}

This function is a smooth approximation of the step-jump in VOF $C$ function. The zero point of the tangent function is shifted (along both  $x$ and $y$ axes) to provide best mapping of the $C$ distribution. The $x_m$ variable signifies the exact $x$ coordinate of tangent function's  shifted zero-point (the ``middle point'' of the THINC-represented interface), and can be calculated using the following:

\begin{equation}\label{th3}
  x_m^{up}=\frac{1}{2\beta}\ln\left( \frac{e^{\frac{\beta}{\gamma}(1+\gamma-2\Phi_{iup})}-1}{1-e^{\frac{\beta}{\gamma}(1-\gamma-2\Phi_{iup})}} \right),
\end{equation}

in which $up$ signifies a properly chosen upwind cell. The flux itself is then available as

\begin{equation}\label{th4}
  f_{i+1/2}=\frac{1}{2}\left( -u_{i+1/2}\Delta t+\frac{\gamma\Delta x}{\beta}\ln\left(\frac{\cosh \lb \beta (\lambda-x_m^{up}-u_{i+1/2}\Delta t / \Delta x \rb}{\cosh \lb \beta (\lambda-x_m^{up}) \rb}\right) \right).
\end{equation}

The parameter $\gamma$ in formulas (\ref{th3}) and (\ref{th4}) controls the slope direction (monotonicity) of the function $\Phi.$ It is chosen to reflect the behaviour of $C$, so

\begin{equation}\label{th5}
  \gamma=
  \begin{cases}
    1 & \Leftrightarrow  C_{i-1}<C_{i+1} \\
    -1 & \Leftrightarrow  C_{i-1}>C_{i+1}.
  \end{cases}
\end{equation}

Since the step center $x_m$ is uniquely determined basing on $C_i,$ the only parameter needed to use formulas (\ref{th3}) and (\ref{th4}) is $\beta,$  the ''slope weighting'' value. Previously, in \cite{xiao} it was constant, but \cite{xiao2011} proposed to set (for $x$ direction)

\begin{equation}\label{th6}
  \beta_x=2.3|n_x|+0.01,
\end{equation}

where $n_x$ is the $x$ component of interface normal $\nb.$ This largely improves the method by dynamically controlling what is called by authors of \cite{xiao2011} ''interface thickness'', that is, the slope used of the hyperbolic tangent function. The role of $\beta$ coefficient is emphasized in \cite{xiao2011}, we have confirmed that even when computing simple droplet translations (both by VOF and CLSVOF approaches), using dynamic $\beta$ greatly improves e.g. interface smoothness. The role and importance of formula choice for (\ref{th6}) is considered widely in \cite{xiao2011}; we will revisit this subject briefly in  \ref{passive}.

The authors of \cite{xiao2011} include in their paper some passive advection tests, such as the Zalesak's problem (rotation of a slotted disc), as well as 3D vortical deformation \cite{menard}, obtaining results comparable to both Youngs \cite{youngs84} and WLIC \cite{yokoi} methods.

Short letter \cite{xiao2011} hasn't introduced any physical flow simulations; still, in a later publication by Li et al. \cite{li2012} some full Navier-Stokes solutions have been presented using the THINC-derived method. However the focus in the latter paper has been on a different, more complicated version - the MTHINC (Multidimensional THINC), employing  the application of surface functions (plane or curved surfaces) that replace in (\ref{th2}) the expression \[\frac{x-x_{i-1/2}}{\Delta x_i}.\] This makes MTHINC relatively complicated, distinguishing it from ``simplified'' approaches, on which we intended to focus here. As such, the MTHINC will not be covered in this article.

\subsection{The Archer CLSVOF method}\label{archsect}


The Archer 3D code, developed at CORIA-INSA institute, is a MAC-type (staggered grid) Navier-Stokes solver. Up to now, it has been applied for example to model primary atomisation \cite{menard, berlemont}, as well as turbulence DNS analysis \cite{duret}. The MGCG (Multigrid with Conjugate Gradients) \cite{brandt,fletcher} methods are used to solve Poisson equation. The Level Set (LS) \cite{osher2000} method is used to track the interface, as well as to couple interface tracking to the pressure solver by the Ghost Fluid (GFM) method \cite{fedkiw-gfm}.

The CLSVOF (Coupled Level Set Volume of Fluid) methodology is used for interface tracking, ensuring traced mass conservation. CLSVOF is a complicated methodology, which is well established and has been comprehensively described both by its creators \cite{sussman2006} as well as in many applications \cite{aniszewski, aniszewskiJCP, menard}.

In short, the method bases upon following concepts. The  Level Set distance function $\phi(\xb),$ defined everywhere in the domain and equal to shortest distance from the interface, can be transported with the flow by direct solving of its transport equation

\begin{equation}
  \label{ls3}
  \frac{D\phi}{Dt}=\frac{\partial\phi}{\partial t}+\mathbf{u}\cdot \nabla\phi=0,
\end{equation}

which is done in case of Archer using the 5-th order WENO\footnote{Weighted Essentially Non-Oscillatory Scheme, \cite{shu}.}. Over time, the $\phi$ function loses its distance property due to numerical error. Hence, a re-distancing technique has been introduced \cite{sussman2006} to improve this, with the re-distance equation

\begin{equation}
  \label{ls4}
  \frac{\partial \phi}{\partial t'}+\mbox{sgn}(\phi_0)(|\nabla\phi|-1)=0,
\end{equation}

solved every few (or every one) iterations of the flow solver. Level Set method alone, especially using GFM \cite{fedkiw-gfm} is a well established tool for interface tracking. It allows for seamless topology changes and facilitates curvature calculation. Additionally, the fact that $\phi$ extends off the interface (is defined everywhere in the domain) has proven useful in applications, such as modelling evaporation \cite{duret}.

The CLSVOF technique assumes, that VOF advection can be performed simultaneously with solution of (\ref{ls3}), and the zero-level set (points with $|\phi|<\epsilon$)  can be corrected to their zero values using VOF interface description. Namely, the $\alpha$ term of equation (\ref{pl0}) is used.

Depending on the approach, a reconstruction of the Level Set represented interface is required in the CLSVOF correction process. To achieve that \textbf{M1} method uses least-square fitting of a plane to the $\phi$ function. In contrast, the \textbf{M4} method uses differential (Youngs' type) expressions to approximate $\phi$ normal vectors. This little difference expresses itself in the fact that the results of \textbf{M1} and \textbf{M4} methods are usually very close, in favor of \textbf{M1} because the least-square technique is potentially of higher order of accuracy than the Youngs' scheme \cite{pilliod}. To couple WLIC and THINC with the level set, an approach identical to \textbf{M4} is used. All tested methods use the same, $\phi$-derived interface normals, and are coupled with Level Set advection in the same way, ensuring a reliable comparison. Effectively, a scenario is possible in which the only difference between the four is narrowed to actual flux computation technique - which is how we have performed all tests described in Section \ref{tests}.

We also note that the distance function $\phi$ is used to set $C=0$ ``far'' from the interface, that is if $|\phi(\xb)|$ grows beyond a set limit in point $\xb$. This $C$ restriction, although used in many published works \cite{sussman2006, aniszewskiJCP} will, in general, break the mass conservation evident in (\ref{sv3}), thus enabling loss of $C$ in CLSVOF approach. More details about the CLSVOF technique can be found in \cite{aniszewski, menard, sussman2006}.

Finally, to facilitate the understanding of how the described methods compare to each other, we have included their features in Table \ref{lstab}. The approximation of $\alpha(C)$ mentioned in Table \ref{lstab} for \textbf{M1} consists in solving (\ref{p31}) with an iterative scheme (akin to Newton-Rhapson) with arbitrary precision. The ``integration'' approach to $F_{x,y,z}(\nb,\alpha,C)$ calculation in $\textbf{M1}$ should be understood as an integration of volume $V(\nb,\alpha)$ for each case of interface position - in which each case is treated separately. This causes immense complication in actual FORTRAN implementation - thus, the $\textbf{M4}$ approach is slightly different, as was described above in context of Figure \ref{plicf2}.

\begin{table}
  \begin{center}
    \caption{Comparative summary of tested methods.}
    \begin{tabular}{|c|c|c|c|c|} \hline
      Method   & \textbf{M1} & \textbf{M2} & \textbf{M3} & \textbf{M4} \\ \hline
      \specialcell{$F_x$ \\  computation} & \specialcell{analytic \\ (integration)} & \specialcell{$\tanh$ \\ approximation} & interpolated & \specialcell{analytic \\ (geometric)}  \\ \hline
      \specialcell{$\nb$ used for \\ $F_{x,y,z}$} & $\nb(\phi)$ &  $\nb(\phi)$ & $\nb(\phi)$ & $\nb(\phi)$ \\ \hline
      \specialcell{$\alpha(C)$ for \\ $F_{x,y,z}$} & approximated & n/a & n/a & analytic \\ \hline
      \specialcell{$\alpha(C)$ in \\ CLSVOF     } & approximated & analytic & analytic & analytic \\ \hline
      \specialcell{Required for \\ VOF advection } & $C,\nb,\alpha$ & \specialcell{$C$ (THINC) or \\ $C,n_x$ (THINC-SW)} & $C,\nb$ & $C,\nb,\alpha$ \\ \hline

    \end{tabular}\label{lstab}
  \end{center}
\end{table}

\section{Numerical Tests}\label{tests}

\subsection{The Flux Curve}\label{fluxcurvsect}

In this subsection, we present what we call the ``flux curve'', that is the plot of the dependence

\begin{equation}\label{fluxcurve}
  F_x\colon = F_x\left(\frac{u\Delta t}{h}\right),
\end{equation}

with $u$ standing for right-hand wall $x$-velocity value $u_{i+1/2,j,k}$, $\Delta t$ being a computed time step, and $h=\Delta x$ a spatial discretisation step.

\begin{figure}[ht!]
  \centering
  \includegraphics[scale=.25]{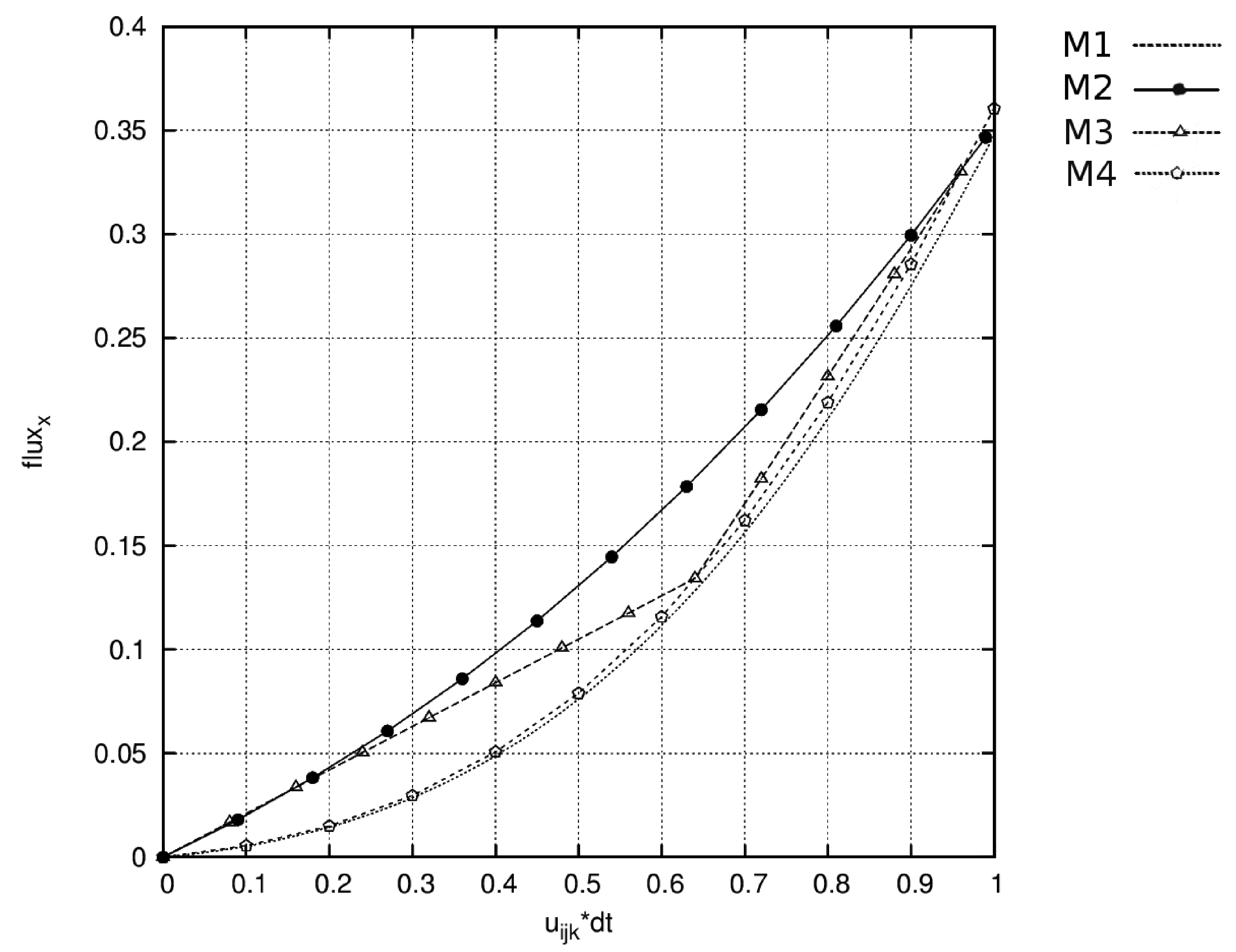}
  \caption{The ``flux curve'' as described in the text.}\label{fc}
\end{figure}

The curves in Figure \ref{fc} have been obtained the following way: a spherical droplet was considered, centered  in a $\lb 0,1 \rb^3$ domain. The radius $r$ of the droplet was set to $r=0.25$ and a grid cell has been selected that contains the point $P=(\frac{1}{2}+\frac{\sqrt{2}}{8},\frac{1}{2}+\frac{\sqrt{2}}{8},\frac{1}{2}+\frac{\sqrt{2}}{8})$. Hence, the point lies on a droplet surface, with all components of $\nb$ positive. Using a $32^3$ uniform grid, we have verified that a grid cell  exists with normalized value of normal components equal to  $\lb \frac{1}{3} ,\frac{1}{3} ,\frac{1}{3} \rb.$  The volume fraction in the cell was $C_{cell}=0.36.$ It was important to create such a configuration to obtain correct $\phi$ values on, and off the interface. Thanks to this, the \textbf{M1} method could have been used.  For this particular $\phi$ and $C$ distributions, both least-squares method (used in \textbf{M1}) and Youngs' procedure (other methods) yields the same values for $\nb$ components.

Using this simple configuration, we have performed sampling of $F_x$ values for varying values of $u_{i+1/2,j,k}$ (the right-hand wall velocity component). The time step $\Delta t$ was set constant. In this configuration, proper mapping of flux values should take the form of $F_x\colon \lb 0,1\rb \rightarrow \lb 0,0.36\rb.$ As we can see from Fig. \ref{fc}, the \textbf{M1} method yields a smooth quadratic curve that changes continuously between points $(0,0)$ and $(1,0.36)$ meaning that $F_x(1)=0.36=C_{cell}$. The difference between the curve obtained for \textbf{M4} (empty circles) and \textbf{M1} is barely visible; still, the methods don't yield identical result, this is caused by differences visible in Table \ref{lstab}.

At the same time, method \textbf{M2} yields a curve which is smooth, but significantly above the \textbf{M1} and \textbf{M4}. The curve for $\textbf{M4}$ is an analytic solution to the $F_x$ problem, hence the overshoot created by \textbf{M2} should be assessed by comparison with \textbf{M4} curve.

As it can be observed in Fig. \ref{fc}, the value for which \textbf{M3} yields an accurate result, is the pivotal value in its definition formulae (\ref{svof1}) and (\ref{svof2}), that is \[1-C_{ijk}=1-0.36=0.64.\] The point
$\left(\frac{u_{i+1/2,j,k}\Delta t}{h}=C_{ijk},F_x(C_{ijk})\right)$ is a common point of two linear segments that ``reconstruct'' the $F_x$ curve in the WLIC method. It is noticeable in Figure \ref{fc}, that for
$\frac{u\Delta t}{h}=C_{ijk}=0.36$ the overshoot (error) in the \textbf{M3} is maximum. At the same time, the \textbf{M2} curve also exhibits a significant error, but the curve's smooth character (see (\ref{th2})) causes the error to change less abruptly with $u\Delta t /h.$ Hence, we conclude that flux values obtained using $\textbf{M3}$ method will be far more \textit{velocity dependent} than for any of the other tested methods, by which we mean that the error produced by it will depend on the local value of velocity used to calculate $F_{x,y,z}$ fluxes.

To conclude this test, let us emphasize that the dependence presented in Figure \ref{fc}  is essential in the assessment of the methods, since it displays, in principle,  all the differences between them in a concise way. However, it is hard to draw conclusions from Fig. \ref{fc}  as to what the actual simulations' results will be. To address this point, we present below the results of various numerical simulations that provide the reader with information about both purely numerical differences between the methods, as well as their influence of the simulated flows' physics.

\subsection{3D passive advections}\label{passive}

\begin{figure}[ht!]
  \centering
  \includegraphics[scale=.60]{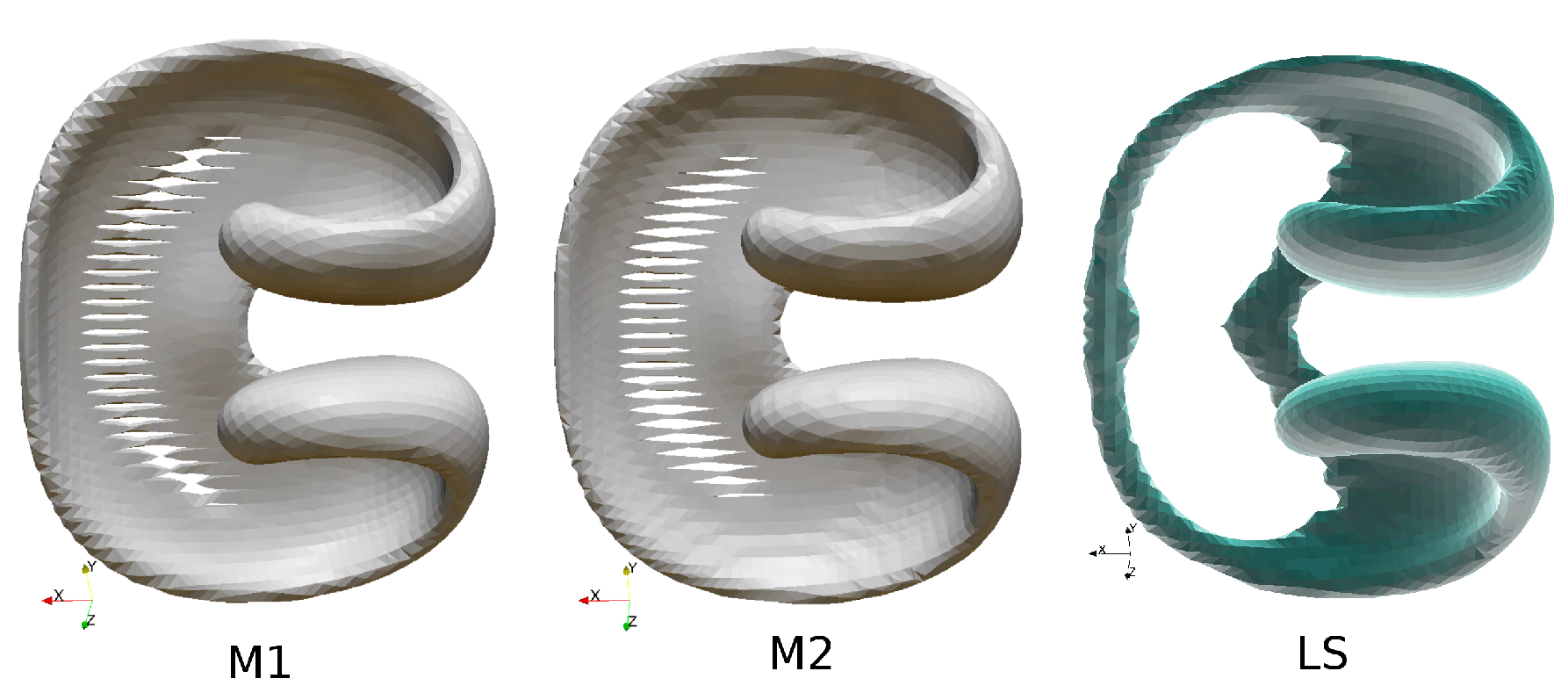}
  \caption{Macroscopic shape of the interfacial surface for the passive advection performed in velocity field (\ref{serp1}), for $t=0.5$s (maximum of deformation). Left: \textbf{M1} method, middle: \textbf{M2} method. Right: advection using only Level Set method (without VOF coupling). }\label{serpentine1}
\end{figure}

To test the behavior of the methods coupled with LS, we have performed first the test of passive advections in the vortical velocity field (\ref{serp1}) defined as

\begin{eqnarray}\label{serp1}
  u(x,y,z,t) & = &2\sin^2\pi x \sin \pi y \sin \pi z \cos \pi t  \\ \notag
  v(x,y,z,t) & = & -\sin\pi x \sin^2 \pi y \sin \pi z \cos \pi t  \\ \notag
  w(x,y,z,t) &=  &-\sin\pi x \sin \pi y \sin^2 \pi z \cos \pi t,
\end{eqnarray}

defined on $\xb \in \lb 0,1\rb$ which has been widely used to test advection methods \cite{menard, xiao, aniszewskiJCP}. A spherical droplet of radius $0.2$ is placed in the point $(0.35,0.35,0.35).$ Field (\ref{serp1}) is usually used to perform advections for $t \in \lb 0,1\rb,$ as it produces the motion of the droplet from the initial position to maximum deformation (see Fig \ref{serpentine1}) and then back to the initial position. If advection is continued for $t>1,$ the droplet undergoes cyclical deformation ad infinitum, which ultimately leads to loss of all traced mass due to numerical error\footnote{\label{mcnote}Note that ''loss of mass'' will effectively mean ''loss of possibility to reconstruct the interface'' (or ''loss of traced volume''), which will be in general a consequence of interface diffusion (''smearing'') and improper calculation of fluxes, especially in split ($x$, $y$ and $z$) approach presented here.}.

In Fig \ref{serpentine1} we present macroscopic shapes of two droplets in velocity field (\ref{serp1}) for $t=1.5 s,$ that is for maximum deformation. Pictured are results of \textbf{M1} (left) and \textbf{M2} (middle) methods using uniform $64^3$ grid. One observes that the result is nearly identical. This legitimate conclusion is shared by Xiao in the work defining the THINC-SW method \cite{xiao2011}; however, further behavior of mass conservation is not studied in \cite{xiao2011} (additionally, a relatively dense $200^3$ grid is used there). We have juxtaposed  these results with ''pure'' Level Set method advection (Fig. \ref{serpentine1}, right-hand side), that is, one in which the distance function $\phi$ is not corrected using VOF $C$ function. It shows clearly that even using a method as simple as \textbf{M2} one is able to significantly improve code's ability to follow a thin film.

\begin{figure}[ht!]
  \centering
  \includegraphics[scale=1.]{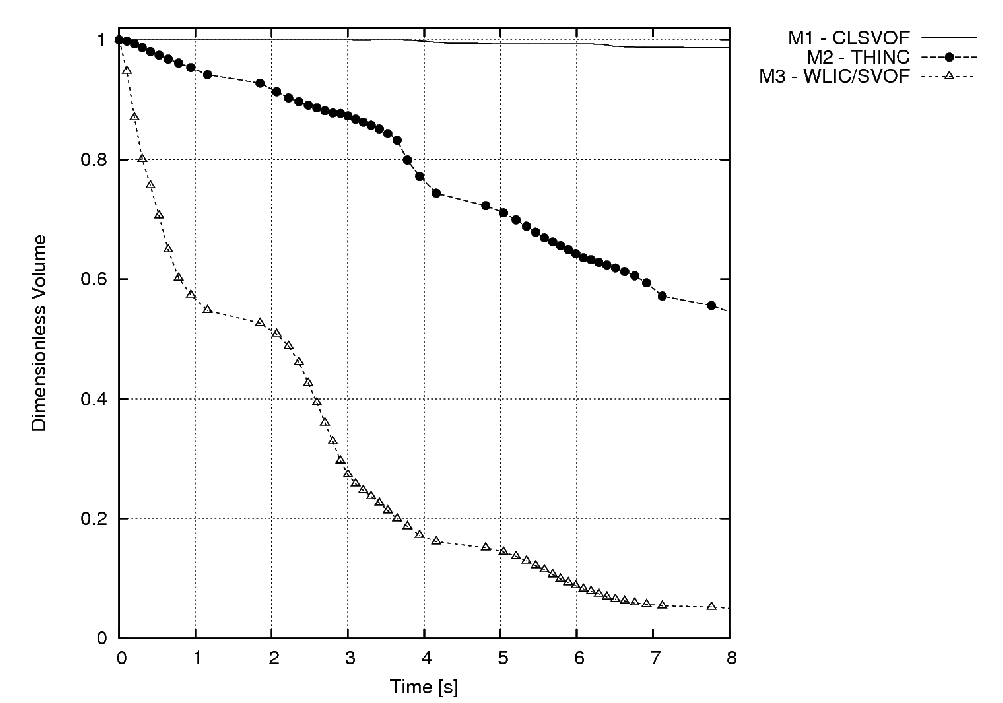}
  \caption{Mass conservation (normalized by the initial mass) for methods \textbf{M1},\textbf{M2} and \textbf{M3}. Results for \textbf{M4} are not shown, since they are identical to \textbf{M1}. }\label{serpentine2}
\end{figure}

Mass conservation becomes an issue when the relatively sparse grid (like uniform $64^3$ grid which was utilized to prepare Figures \ref{serpentine1} and \ref{serpentine2}) is used, and when considering larger time interval $1s<t<8s.$ In Figure \ref{serpentine2}, clear loss of traced mass is observable for the \textbf{M2} method, compared to \textbf{M1}. After $8$s of simulated time, nearly half of the traced mass is loss.

The mass conservation of the $\textbf{M3}$ method falls even below that of $\textbf{M2}$, with  all mass lost on grids more sparse than $64^3$ (see Table \ref{convtable}). However, the macroscopic shape of the droplet for $t=0.5$s using M3 and $128^3$ grid is identical to those presented in Fig \ref{serpentine1}. Also, Marek et al. \cite{svof} (see Figure 10 therein) show in their implementation of the method that mass conservation at the level of $90 \%$ should be expected for $t\approx 0.5s.$  This is consistent with results presented here, although the WLIC method was used in their work without coupling it to LS, and finally,  the grid $150^3$ was used in \cite{svof}.

\begin{table}
  \begin{center}
    \caption{The $L_1$ error in passive advection using (\ref{serp1}) and four tested methods (coupled with Level Set method).}

    \textbf{M1}

    \begin{tabular}{|c|l|c|c|}
      \hline
      grid & $L_1$ error & rate & order \\
      \hline
      16$^3$  & 1.826$\times 10^{-2}$ & --     & -- \\
      32$^3$  & 8.928$\times 10^{-3}$ & 2.044  & 1.02\\
      64$^3$  & 6.771$\times 10^{-3}$ & 1.318  & 0.659 \\
      128$^3$ & 3.256$\times 10^{-4}$ & 7.250  & 3.62\\
      \hline
    \end{tabular}
    
    \textbf{M2} 

    \begin{tabular}{|c|l|c|c|}
      \hline
      grid & $L_1$ error & rate & order \\
      \hline
      16$^3$  & 2.119$\times 10^{-2}$ & -- & -- \\
      32$^3$  & 1.029$\times 10^{-2}$ &  2.059  & 1.02\\
      64$^3$  & 5.175$\times 10^{-3}$ & 1.988 & 0.99 \\
      128$^3$ &  1.808$\times 10^{-3}$  &  2.862  & 1.43\\
      \hline
    \end{tabular}

    \textbf{M3}$^a$

    \begin{tabular}{|c|l|c|c|}
      \hline
      grid & $L_1$ error & rate & order \\
      \hline
      16$^3$  & n/a & -- & -- \\
      32$^3$  & n/a &  n/a  & n/a\\
      64$^3$  & 2.537$\times 10^{-2}$ & n/a & n/a \\
      128$^3$ &  1.116$\times 10^{-2}$  &  2.273  & 1.13\\
      \hline
    \end{tabular}

    \textbf{M4}

    \begin{tabular}{|c|l|c|c|}
      \hline
      grid & $L_1$ error & rate & order \\
      \hline
      16$^3$  & 1.755$\times 10^{-2}$ & -- & -- \\
      32$^3$  & 8.424$\times 10^{-3}$ &  2.083  & 1.04\\
      64$^3$  & 6.720$\times 10^{-3}$ & 1.253 & 0.62 \\
      128$^3$ & 7.583$\times 10^{-4}$ &  8.861  & 4.43\\
      \hline
    \end{tabular}
    \label{convtable}

    $^a$\footnotesize{All mass lost on grids 16$^3$ and 32$^3$.}
  \end{center}
\end{table}

The passive advection using (\ref{serp1}) is an appropriate test-case to yield method convergence. As shown in Table \ref{convtable}, we have performed in total $16$ simulations with varying grid size, and measured the $L_1$ error of advection by means of

\begin{equation}
\label{l1error}
L_1=\frac{1}{N^3}\left(\sum\limits_{i,j,l=1}^N |C^0_{ijk}-C^f_{ijk}|\right),
\end{equation}

where $N$ is a grid size (uniform in all directions). Symbols $C^0$ and $C^f$ are used for $C$ distributions at $t=0$ and $t=1,$ respectively. The ''rate'' column shows approximate error decrease, also an order estimation is included. The \textbf{M4} method produces smallest errors, while \textbf{M3} performs worst, causing the loss of all traced mass for grids $16^3$ and $32^3.$ We observe, that while the order of all methods is in most cases approximately $1,$ the \textbf{M1} and \textbf{M4} produce substantial error decrease when the grid size changes from $64^3$ to $128^3,$ an effect absent while using simplified VOF approaches.

\begin{figure}[ht!]
  \centering           
  \includegraphics[scale=.73,angle=0]{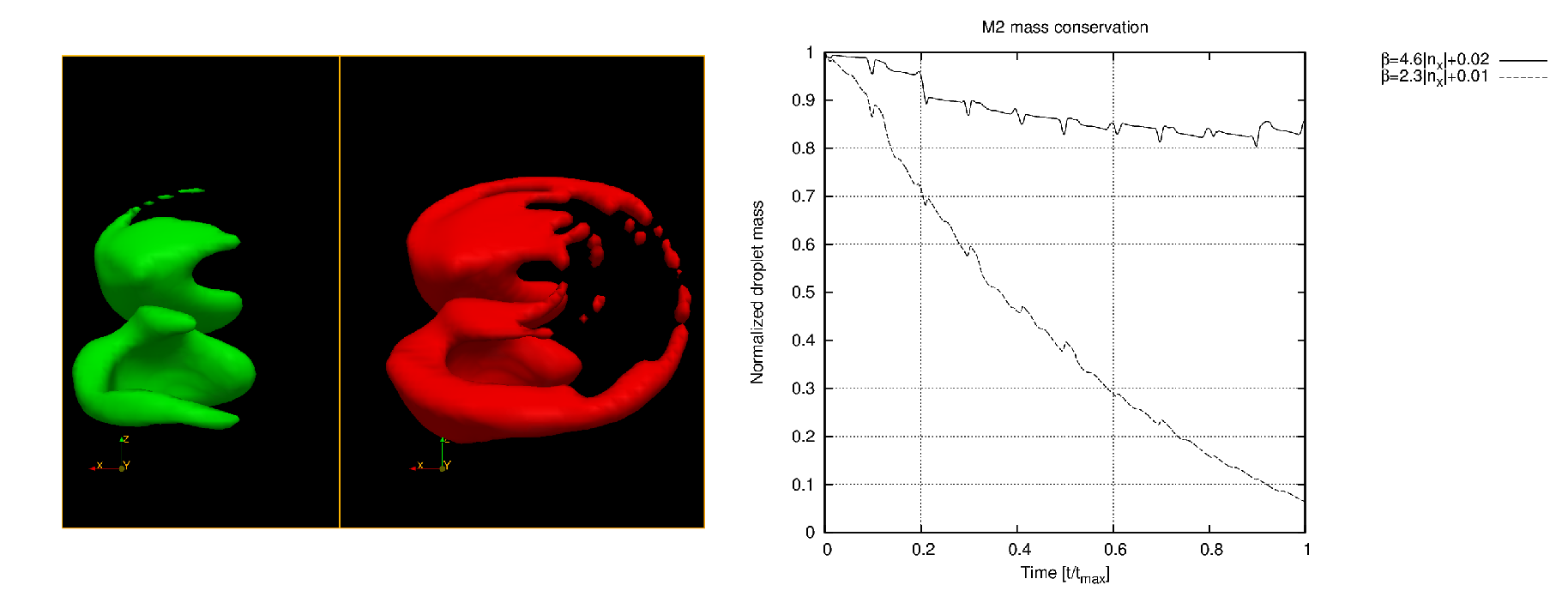}
  \caption{Left: Passive advections with (\ref{serp1}); results using \textbf{M2} method with $\beta_x=2.3|n_x|+0.01$ (green, leftmost image) and using $\beta_x=4.6|n_x|+0.02$. Right: Normalized mass conservation as a function of $t/t_{max}$ for the same two simulations. }\label{thincvariants}
\end{figure}

Before concluding the description of the passive advection using velocity field (\ref{serp1}), we would like to add a remark concerning method \textbf{M2}. When describing formula (\ref{th6}), we mentioned additional possibilities in approaching ''interface thickness'' coefficient \cite{xiao2011}. To emphasize the importance that $\beta$ has for mass conservation, let us examine Figure \ref{thincvariants}, which displays the impact of changing $\beta$ from (\ref{th6}) to 

\begin{equation}\label{th6alt}
\beta_x=4.6|n_x|+0.02,
\end{equation}

i.e. multiplying it by two. As one can easily observe, the mass conservation was drastically improved by the change to (\ref{th6alt}). This is caused by a decrease of spatial spread in $C$ fractal function, that is to say, decrease of ``interface thickness'' in the sense used in \cite{xiao2011} which is the number of grid meshes with nontrivial $C$ values ($0<C<1$) in the vicinity of interface. In PLIC methods, such as \textbf{M4} this spread doesn't take place, since, if we cross the interface along its normal direction, we should always encounter only one nontrivial cell. In THINC/SW method, however, the decision of what is to be considered an actual interface location is arbitrary -- one could settle for $C=0.5$ as do we, but different values could be considered \cite{xiao, xiao2011}, the more so  since the sum 

\begin{equation}\label{vofsum}
\sum\limits_{i,j,k} C_{i,j,k}
\end{equation}

is conserved as a consequence of (\ref{sv3}). 

By changing $\beta$ formulations from (\ref{th6}) to (\ref{th6alt}), we have effectively decreased the number of nontrivial cells, thus the number of cells with $C>0.5$ has increased as well, which is the reason for change visible in Fig. \ref{thincvariants}. All VOF-type methods exactly conserve the $C$ function since (\ref{vofsum}) is constant between $C^{n}$ and $C^{n+1};$ however, the reconstructed \textit{volume} (one could roughly think of a volume delimited by cells having a given $C$ value) will generally not be conserved when the interface has nonzero ''thickness''.  Additionally, it poses difficulties for CLSVOF implementation, as explicit, reconstructed  interface location is needed to compare $C$ with $\phi$ (see also footnote \ref{mcnote}). Changing the \textbf{M2} ''interface thickness'' coefficient to (\ref{th6alt}) for following numerical tests was not considered, since firstly, the publication \cite{xiao2011} favoured (\ref{th6}), and secondly, numerical tests suggested that it might influence the smoothness of the reconstructed interface. 

\subsection{Static droplet test}

In this subsection, we describe the static droplet test. Performed for example as a means to test curvature calculation routines \cite{popinet1, popinet2} the test includes a static, spherical droplet suspended in a periodic domain without gravity. Full Navier-Stokes equations are solved, and surface tension is taken into account. Density ratio of ``liquid'' and ``gas'' is kept unitary, that is $\rho_g=\rho_l=1000.$ The domain size is set to $L_x=0.1$, while droplet diameter $d=0.5L,$ with the droplet being placed in domain's center.  Simulated time $t_{max}=5$, is used together with constant, non-zero viscosity  $\mu_g=\mu_l=0.03$ and a surface tension coefficient $\sigma=0.045.$ One expects no motion of the interface, since the system is in equilibrium. However, numerically, the curvature calculation scheme might result in non-symmetric pressure distribution, and though the interaction with $\ub$ procure its fluctuations popularly known as ''spurious currents''.

When $\kappa$ is imposed (fixed to $2/d$ on the interface) the system remains static, and $\max \ub$ is $0$ to machine precision\footnote{Which we have tested while pre-setting the calculation proper.}. Obtaining $\kappa$ via numerical method such as CLSVOF will result in ''parasitic currents'' which in turn will reshape the interface (here, advection using \textbf{M1-4} methods takes place), creating a series of oscillations that may be dumped numerically \cite{popinet1}. Such a system is deemed useful by us to amplify advection errors, since those should result in further breakdown of initial droplet symmetry and be observable in $\max \ub.$  In all cases, actual curvature is calculated using $\kappa(\phi)=\nabla\cdot\nb(\phi),$ implemented in an efficient way\cite{menard}. We only intend to check if the choice of coupled  VOF method will influence the $\phi$ distribution in a manner that changes the flow character.

\begin{figure}[ht!]
  \centering
  \includegraphics[scale=1.]{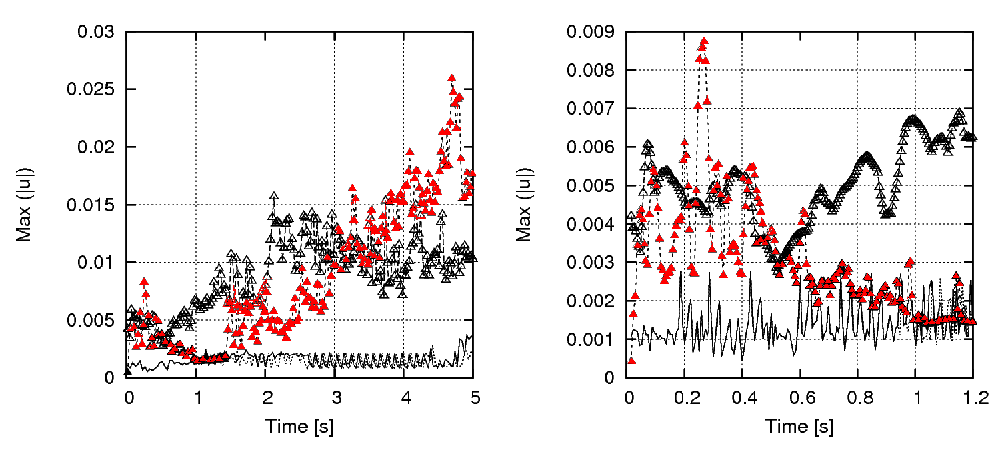}
  \caption{Maximum velocity norm $|\ub |$ observed for the static droplet  case using a $32^3$ grid; plotted for $t\in \lb 0,5\rb$ (left) and $t\in \lb 0,1.2 \rb.$ \textbf{M1} - continuous line, \textbf{M2} - empty triangles, \textbf{M3} - red triangles, \textbf{M4} - dashed line.}\label{32oscill}
\end{figure}

In Figure \ref{32oscill} one can observe the behaviour of the maximum norm of $\ub$ in the simulation performed using the $32^3$ rectangular grid.


In this kind of flow, one expects a series of oscillations dissipating energy due to the action of both physical viscosity and what is called ''numerical viscosity''. Since $\mu_l\ne 0 \wedge \mu_g\ne 0,$ the series of oscillations is expected to be quickly extinguished. Indeed, this can be observed by following continuous lines in Figure \ref{32oscill} signifying $\max|\ub |$ for \textbf{M1} method. At the same time, the \textbf{M2} (THINC) method (empty triangles) produces bounded (in $t \in \lb 0,5\rb$) series of oscillations, which converge approximately at $\max(|\ub|)=0.1.$ The \textbf{M3} (WLIC/SVOF) method (red triangles) yields  values of order similar to  \textbf{M2}, we also  notice some movement of the droplet along the $z$ axis  from it's initial position. However, in tested time interval $t\in \lb 0,5\rb$s, using \textbf{M3} doesn't lead to droplet disintegration, and its shape remains spherical. We see that for \textbf{M1} and \textbf{M4} methods, maximum velocity value is of the order of $0.004,$ while both \textbf{M2} and \textbf{M3} methods center around much higher values, $0.01$ and $0.02$ respectively. Also, in Fig. \ref{32oscill} we can see that in time interval $t\in\lb 0,1\rb$ the \textbf{M1} and \textbf{M4} methods yield exactly the same results. This might suggest that the onset of oscillation mechanism, in its beginning phase, not fed by the numerical error in $F_{x,y,z}$ calculation, but instead by factors such as imbalance in (discrete) pressure jump on both sides of the droplet. For the $t>1$s however, the flux calculation errors and shape distortions resulting from them seem to become important, causing the growth of oscillations in the case of \textbf{M2} and \textbf{M3} methods.

Considering \textbf{M3} method, the tests presented in Fig. \ref{32oscill} were made using Variant III from Table \ref{svof22}, since it produced smallest ''parasitic''  currents (maxima of $|\ub |$), using all of the tested grids. Different variants led to non-convergence and droplet disintegration.

\begin{figure}[ht!]
  \centering
  \includegraphics[scale=1.]{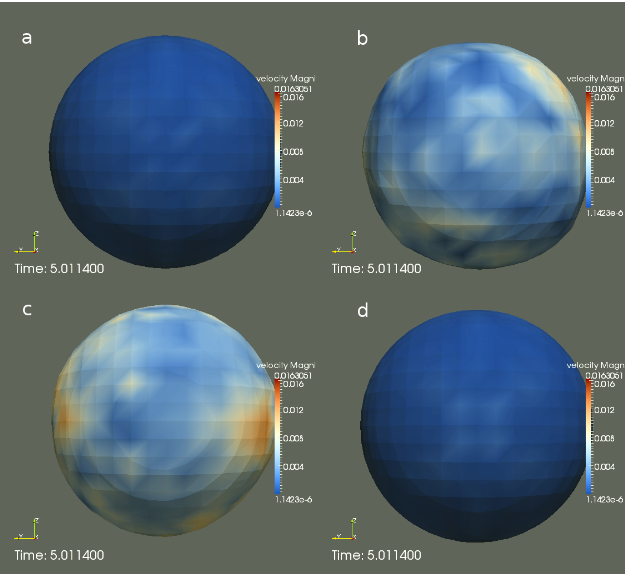}
  \caption{Droplet shapes obtained using a $32^3$ grid for $t=5$s at the end of static droplet test simulations. Images for \textbf{M1} (a), \textbf{M2} (b),\textbf{M3} (c),\textbf{M4} (d), coloured by $|\ub|.$ }
\end{figure}

In \cite{yokoi2013} K. Yokoi has used the WLIC method to simulate the rising bubbles and splashes. Therefore, it would be reasonable to expect the existence of ``spurious currents'' effect could manifest themselves in this work. However, \cite{yokoi2013} employs a CSF (Continuum Surface Force of Brackbill et al. \cite{brackbill}) scheme to calculate surface tension, which is not directly comparable with LS-based $\kappa$ calculation and Ghost-Fluid techniques used in this work (moreover, CSF schemes are known to cause smoothing of the interface which could counteract errors caused by the interface tracking). Also, dense grids have been used (e.g. $256\times 256\times 64$ to simulate the splash of a single droplet), hence the curvature was very well resolved.

\begin{table}
  \begin{center}
    \caption{Static droplet test: surface tension calculation errors  for four tested methods using Ghost Fluid Method and $\phi$-based curvature calculation.}
    \begin{tabular}{l|c|c|c}
      Grid size  & $L_1$   & $L_2$   & $|u_{max,50}|$ \\      
      \cline{1-4}
      \textbf{M1/M4} &        &        &       \\
      $16^3$    &   $1.203\times 10^{-3}$      &   $4.861\times 10^{-3}$     &    $0.293\times 10^{-4}$    \\
      $32^3$    &   $1.160\times 10^{-3}$     &    $4.775\times 10^{-3}$    &    $0.761\times 10^{-4}$    \\
      $64^3$    &    $1.228\times 10^{-3}$     &    $4.912\times 10^{-3}$    &     $0.680\times 10^{-2}$   \\
      $128^3$    &   $1.185\times 10^{-3}$      &   $4.820\times 10^{-3}$     &    $0.826\times 10^{-5}$    \\
      \cline{1-4}
      \textbf{M2} &        &        &       \\
      $16^3$    &   $1.205\times 10^{-3}$      &   $4.868\times 10^{-3}$     &    $0.858\times 10^{-2}$    \\
      $32^3$    &   $1.170\times 10^{-3}$     &    $4.816\times 10^{-3}$    &    $0.443\times 10^{-2}$    \\
      $64^3$    &    $1.229\times 10^{-3}$     &    $4.921\times 10^{-3}$    &     $0.605\times 10^{-2}$   \\
      $128^3$    &   $1.237\times 10^{-3}$      &   $5.063\times 10^{-3}$     &    $0.106\times 10^{-1}$    \\
      \cline{1-4}
      \textbf{M3} &        &        &       \\
      $16^3$    &   $1.203\times 10^{-3}$      &   $4.860\times 10^{-3}$     &    $0.212\times 10^{-4}$    \\
      $32^3$    &   $1.160\times 10^{-3}$     &    $4.773\times 10^{-3}$    &    $0.428\times 10^{-2}$    \\
      $64^3$    &    $1.221\times 10^{-3}$     &    $4.890\times 10^{-3}$    &     $0.677\times 10^{-2}$   \\
      $128^3$    &   $1.230\times 10^{-3}$      &   $5.003\times 10^{-3}$     &    $0.128\times 10^{-1}$    \\
      \cline{1-4}
    \end{tabular}\label{statictab}
  \end{center}
\end{table}

Alternatively, it is possible to use the same static droplet test setup to assess the quality of surface tension calculation scheme with four tested methods, by comparing the pressure field with Laplace pressure. Although GFM and $\phi$-derived curvature is used in all cases, the combinations with \textbf{M1}-\textbf{M4} methods may in general yield different results when more than single time-step is considered, i.e. when advection-induced errors might appear. Following Gerlach et. \cite{gerlach2005} we are  comparing the Laplace pressure inside the spherical droplet 

\begin{equation}
P_d = \sigma \kappa = 2 \frac{\sigma}{R}\label{lppr}
\end{equation}

to  values obtained from simulation, by the means of $L_1$ and $L_2a$ errors computed as follows:

\begin{equation}\label{l1stat}
L_1=\left| \frac{\Sigma_{ijk}^{N_d}P_{ijk}-Pd}{N_{d}P_d}\right|,
\end{equation}
\begin{equation}\label{l2stat}
L_2=\left\lb \frac{\Sigma_{ijk}^{N_d}(P_{ijk}-Pd)^2}{N_{d}P_d^2}\right\rb^\frac{1}{2}.
\end{equation}

where $N_d$ is the number of cells in the interior of the droplet ($C\ge0.99$).  Since for initial simulation time-step (with $\ub=0$ everywhere) all the methods are bound to yield the same result, we have measured $L_1$ and $L_2$ after $50+1$ solver iterations using four grid sizes. The timestep is fixed to $\Delta t\approx 0.79\times 10^{-3},$ leading to $t_{end}\approx 0.408\times 10^{-1}.$ This gives a set of $16$ simulations, whose results we present in Table \ref{statictab}. Additionally, maximum velocity norm $|u_{max,50}|$ is presented (thus, for each method the row in Table \ref{statictab} corresponding to the $32^3$ grid has a value belonging to set of points plotted in Fig. \ref{32oscill}). There is a common row in Table \ref{statictab} for \textbf{M1/4} methods, since the results were identical within given accuracy.

Concerning $|u_{max,50}|,$ the behaviour of \textbf{M1/4} is consistent with results of Gerlach et al. \cite{gerlach2005} who found that the convergence of CLSVOF is flawed in this aspect; we note however the a sharp error decrease when $128^3$ grid is used, which was not reported in \cite{gerlach2005}.  The simplified methods \textbf{M2} and \textbf{M3} produce $|u_{max,50}|$ results much inferior to \textbf{M1/4} which, consistently with Fig. \ref{32oscill} proves that ``parasitic currents'' phenomenon is present, and is not directly remedied by increasing resolution in this setup.

It is interesting to note that all four methods suffer from lack of convergence in curvature/surface tension calculation. This is a well documented behaviour of VOF-based schemes, described e.g. by Raessi et al\cite{cummins}  or Cummins et al. \cite{Raessi2} which does not appear when using ''pure'' Level Set method for the same setup, meaning that the lack of satisfactory convergence is a result of $\phi$ being corrected using $C$ function. Authors of \cite{Raessi2} have found the same behaviour when examining $\phi$-derived curvatures for a circle with grids ranging from $16^3$ to $1024^3,$; the CLSVOF method which uses $\kappa  \sim \nabla \cdot \nabla\phi$ introduces an zeroth-order error, which is ``always associated with $\kappa$ regardless of the mesh resolution''.

In case of study presented here, Table \ref{statictab} shows that for all methods the $L_{1,2}$ errors are minimal when the $32^3$ grid is used. Other from that, the errors remain nearly static when the grid is changed, which seems consistent with aforementioned \cite{Raessi2} result.   It is therefore hard to make comparative assessments; however it is visible that when considering $L_1$ and $L_2$ errors, the \textbf{M3} slightly outperforms  \textbf{M2}. Also, when considering these errors' orders of magnitude,  \textbf{M2} and \textbf{M3} may be seen as comparable with \textbf{M1/4}. Thus, as far as balancing the pressure field by surface tension force is considered, we conclude that  all the tested methods yield equivalent results.

It seems  encouraging then to test how the ''parasitic currents'' phenomenon described above,  and possible errors induced by advection into the  $\kappa$ calculation will influence a surface-tension driven phenomenon more dynamic in character.

\subsection{Oscillating Droplet} \label{droposc}

The brief ''parasitic currents'' study presented above may be seen as non-conclusive if one is to assess the behaviour of \textbf{M1-4} methods in surface tension dominated/driven flows. Thus, we follow with simulations set  addressing the oscillation of a liquid droplet -- a phenomenon which has been subject to extensive linear analysis by Lamb \cite{lamb} -- who used spherical harmonics to produce formulas for oscillation frequencies of individual modes -- and numerous computational studies both in two and three dimensions, such as the ones by Basaran \cite{basaran} or Shin \& Juric \cite{shin_juric}. 

We start by setting up a problem geometry virtually identical to Lamb's description (see \cite{lamb} p. 469) of example water droplet that ``vibrates seconds'', that is, takes one second to finish prolate/oblate cycle. Hence, a water (''Fluid 2'') droplet of radius $r=2.47$cm is simulated in a cubical domain of size $L=4\cdot r.$ Liquid properties correspond to water at room temperature i.e. $\rho_2=1000$ (kg/m$^3$), $\mu_2=1.79\cdot 10^{-3}$(Pa$\cdot$s), $\sigma=0.074$ (N/m). The droplet is suspended in lighter fluid (``air'' or ''Fluid 1'') - for Lamb's analysis to hold, we should provide $\rho_1<<\rho_2,$ so that lighter fluid is neglected. We chose the light fluid to have the viscosity of air (so $\mu_1=0.017$ (Pa$\cdot$s)) however, we kept its density at $\rho_1=10$ for numerical reasons\footnote{Avoiding the decoupling between numerical mass and momentum transfers, as described by Raessi \& Pitsch \cite{Raessi}.}, set the droplet centered in the domain with initial elongation of its $z$ radius  $ r_z=  1.04 \cdot r,$ and observed its oscillations (using various grids) over $3.5$ seconds. 

\begin{figure}[ht!]
  \centering
  \includegraphics[scale=.99]{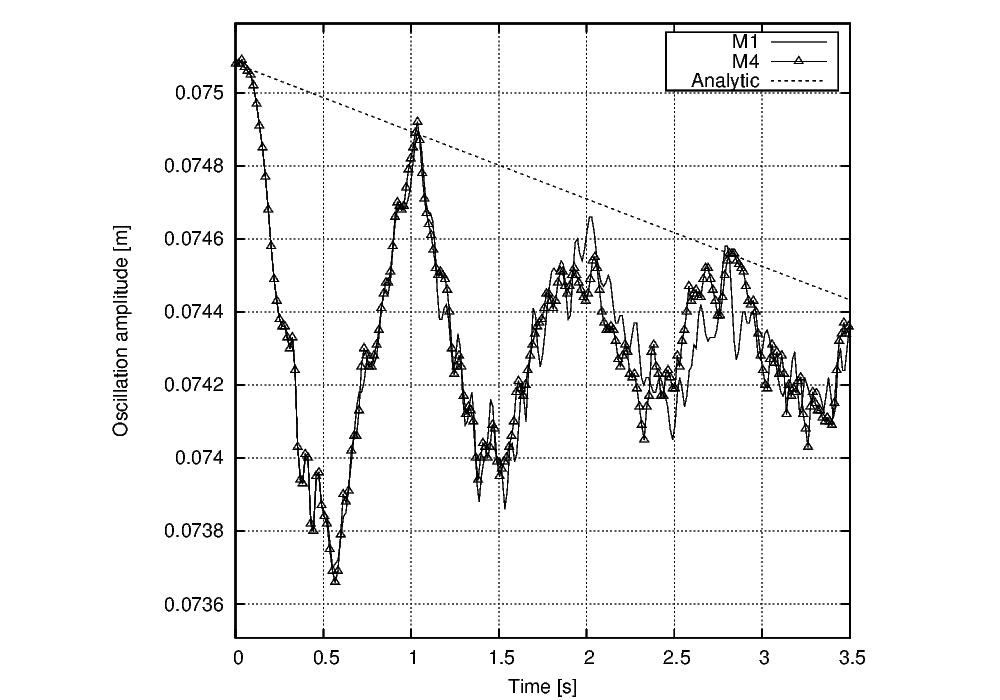}
  \caption{Oscillation amplitude using \textbf{M1} and \textbf{M4} methods, with a fit produced using (\ref{osc4}). }\label{oscillation1}
\end{figure}

Firstly, Lambs prediction of $\omega_2,$  the frequency of second mode, was examined. For the $n-$th mode, Lamb \cite{lamb,shin_juric} found that the frequency can be expressed as

\begin{equation}\label{osc2}
\omega_n=\frac{n(n+1)(n-1)(n+2)\sigma}{\lbrack(n+1)\rho_1+n\rho_2\rbrack r^3},
\end{equation}

leading to the conclusion that for an oscillating droplet of water, the dominating second mode has the frequency

\begin{equation}\label{osc3}
\frac{\omega}{2\pi}\approx 3.87 r^{-3/2},
\end{equation}

which amounts to $1-$ second long intervals when $r=2.47$ cm. This behaviour can be seen in Fig. \ref{oscillation1}, which displays a complete temporal evolution of amplitude simulated using $32^3$ grid with \textbf{M1} and \textbf{M4} methods. We can see that although both methods' predictions differ for $t>2$s, the frequencies are well resolved, in particular the first oscillation occurs nearly at $t=1$s. Quantitative assessment is found in Table \ref{oscillationtab1}, which contains errors calculated as $|\omega_{sim}-\omega_2|$ for all tested methods. Additionally, length of second oscillation is presented in Table \ref{oscillationtab1}.

A curve visible in Figure \ref{oscillation1} is the amplitude decay of the $n$-th mode \cite{shin_juric}, given by

\begin{equation}\label{osc4}
a_n(t)=a_0 e^{-t/\tau_n},
\end{equation}

with

\begin{equation}\label{osc5}
\tau_n = \frac{r^2}{(n-1)(2n+1)\nu}
\end{equation}

where $\nu$ is the kinematic viscosity of fluid $2$. We can see that the \textbf{M1} and \textbf{M4} decay rates fit closely the $a_2(t)$ curve, amounting to $0.042\%$ and $0.12\%$ of averaged error respectively. These values, together with Table \ref{oscillationtab1} show overall good agreement of \textbf{M1}/\textbf{M4} predictions' with analytic results compared to previous publications \cite{shin_juric, basaran}, in spite of having used a coarse grid.

\begin{table}
  \begin{center}
    \caption{Error (absolute) in prediction of $\omega_2$ frequency (measured in two initial cycles of oscillation) using $32^3$ grid.}
    \begin{tabular}{|c|c|c|c|c|}
      \cline{1-5}
              & \textbf{M1} & \textbf{M2} & \textbf{M3} & \textbf{M4} \\
      \cline{1-5}
      cycle 1 &    $2\%$     &    $30\%$    &     $9\%$    &     $2\%$    \\
      cycle 2 &    $1\%$     &    N/A      &     $10\%$    &     $1\%$    \\
      \cline{1-5}
    \end{tabular}\label{oscillationtab1}
  \end{center}
\end{table}

\begin{figure}[ht!]
  \centering
  \includegraphics[scale=1]{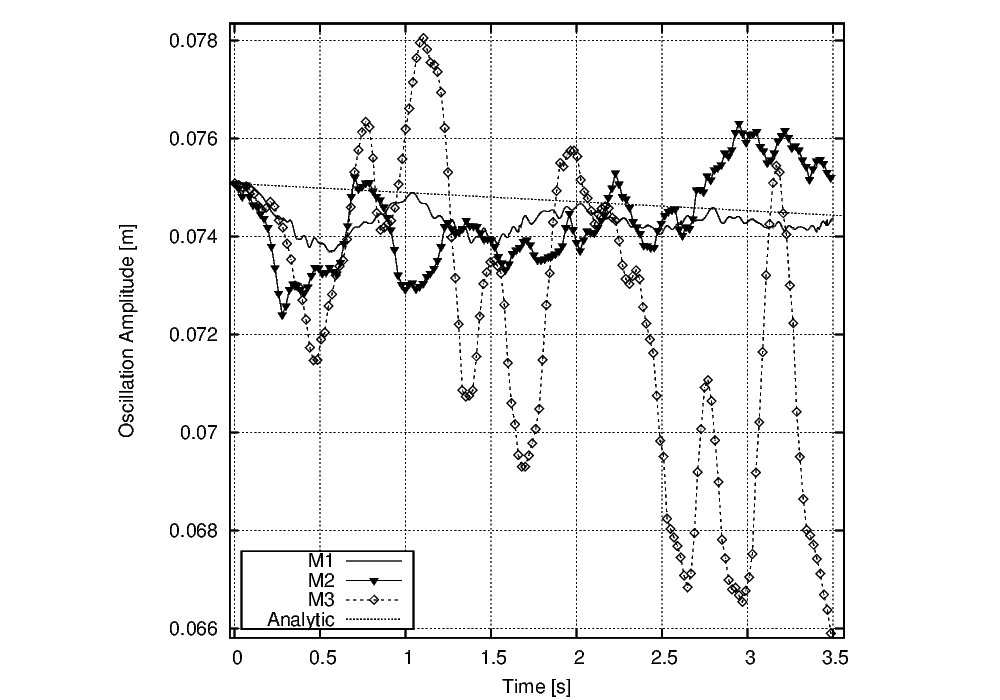}
  \caption{Oscillation amplitude  using \textbf{M1},\textbf{M2} and \textbf{M3} methods, with  (\ref{osc4}) decay curve. Method \textbf{M4} has been omitted, since at this scale the plot seems identical to \textbf{M1}. }\label{oscillation2}
\end{figure}

Both \textbf{M2} and \textbf{M3} methods perform unsatisfactorily when simulating this surface-tension driven phenomenon. Inspecting Figure \ref{oscillation2} we can judge that the first oscillation period of $1$s is mispredicted by \textbf{M3} ($9\%$ error, Table \ref{oscillationtab1}), while \textbf{M2} produces significant $30\%$ error - the droplet ends its first oscillation cycle after $0.70$s. These errors grow rapidly for both ''simplified'' methods as the simulation progresses. 

\begin{figure}[ht!]
  \centering
  \includegraphics[scale=1.]{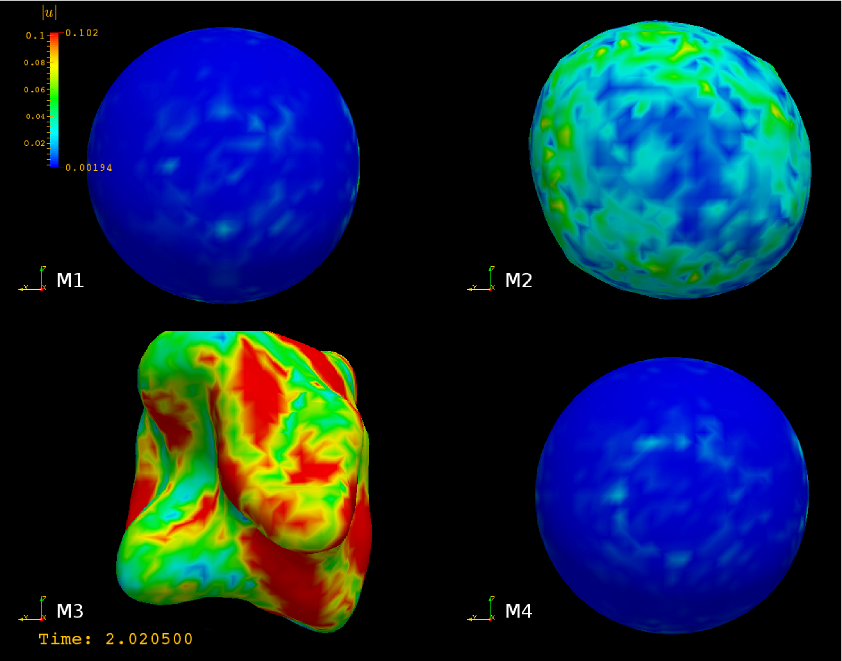}
  \caption{Simulated droplet shapes for oscillating droplet test case, $t\approx 2.02$s, using $64^3$ grid and all tested method, coloured by $|\ub|.$ }\label{oscillation3}
\end{figure}

Figure \ref{oscillation2} shows that for \textbf{M2} the oscilation amplitude des not decay. The situation is even worse for \textbf{M3}: growth of amplitude is uncontrolled (one order of magnitude over analytic prediction), which eventually breaches the spherical oscillation regime to which Lamb's description applies. This can be further verified by inspecting Figure \ref{oscillation3}, in which oscillation shapes are presented for $t\approx 2$s. The distorted droplet shape produced by \textbf{M3} is clearly visible; moreover, as we have verified using $32^3$ and $64^3$ grids, the droplet is later destroyed. 

This lack of stability when simulating oscillations of droplet using \textbf{M3} might be attributed to minute errors introdced by WLIC-derived fluxes of $C$ to interfacial  curvature, that in turn could be amplified when using Ghost Fluid Method, as is the case in our work. Thus, we don't rule out that when using different surface tension calculation schemes (e.g. CSF-type LS or VOF based methods with smoothing \cite{yokoi2013}) could ballance this phenomenon. 

\subsection{Phase Separation Simulation}

\begin{figure}[ht!]
  \centering
  \includegraphics[scale=0.4]{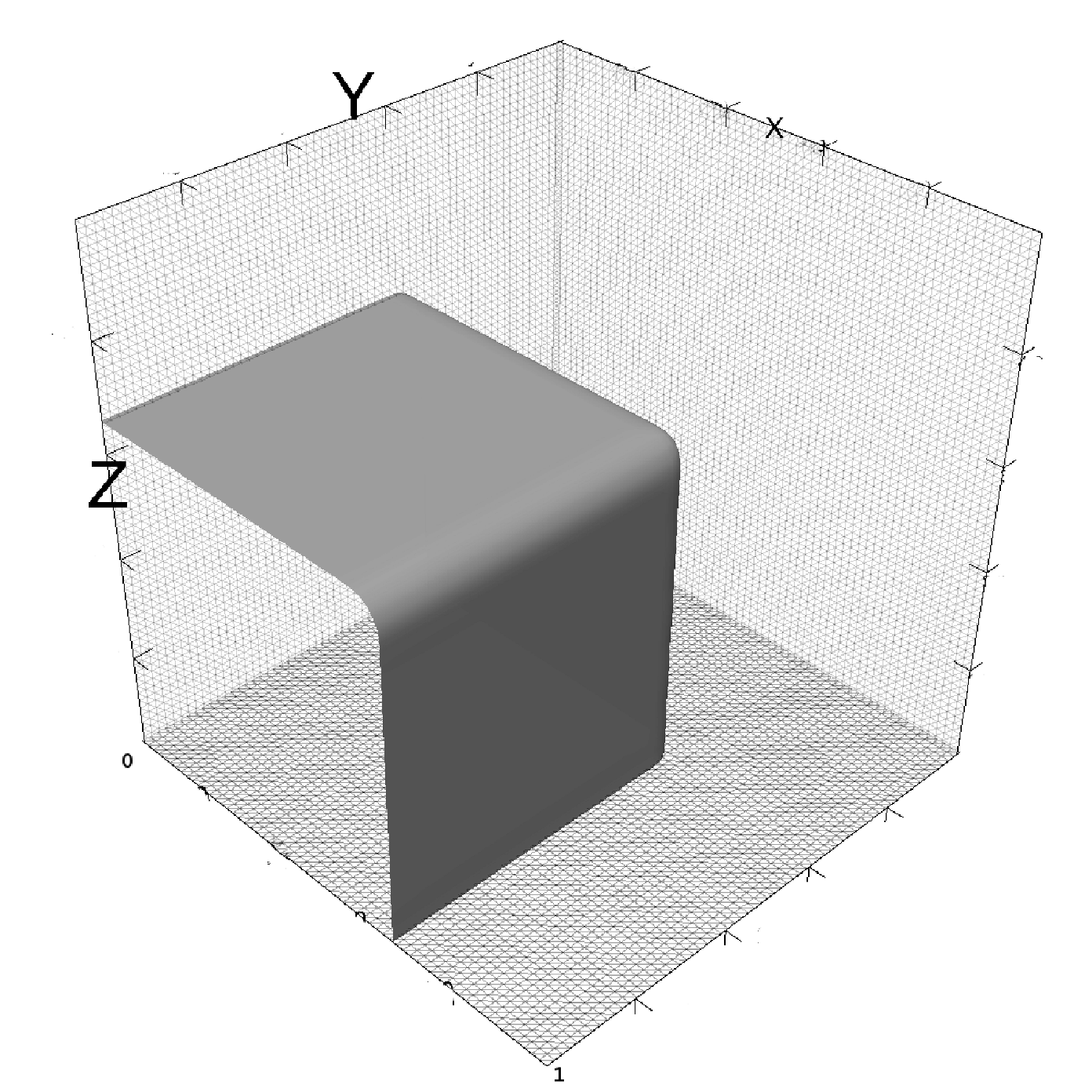}
  \caption{Position of the interface for $t=0$ in the phase separation simulation.}\label{mix0}
\end{figure}

The phase separation setup includes a cubical box of side length $L=1$ m , a lighter fluid (``oil'') with $\rho_o=900$ kg.m$^{-3}$  placed in octant closest to $(0,0,0)$ point. Heavier fluid (``water'') with density  1000 kg.m$^{-3}$ occupies the rest of the box. See Fig. \ref{mix0} for reference (after \cite{aniszewski}).  The viscosities of water and oil are respectively $0.001$ and $0.1$ Pa$\cdot$s, and surface tension is set to 0.075 N$\cdot$m$^{-1}$.

This numerical setup has been used in \cite{vincent, larocque} to investigate sub-grid terms in a turbulent two-phase flow, and in a work by one of us (Aniszewski) \cite{aniszewskiJCP} which suggested a model for sub-grid surface tension. The test case is characterized by high Reynolds numbers \cite{vincent} (up to $7\cdot 10^5$) and a developed interfacial are, with strong generation of small structures. In this short paper, we present the simulations presented on a $128^3,$ uniformly spaced grid. Authors of \cite{vincent} conclude that this grid is not sufficient to resolve all the scales of the flow. This authors' results obtained on a $256^3$ and $512^3$ grid are at the moments yet unpublished, however, we have reasons to believe that neither one of that grids provides what could be considered ``DNS'' (Direct Numerical Simulation) of this flow\footnote{Since we have included no sub-grid models in the simulations presented here, they should be seen as the so-called ``Implicit Large Eddy Simulations'' \cite{domaradzki}}.

\begin{figure}[ht!]
  \centering
  \includegraphics[scale=1.0]{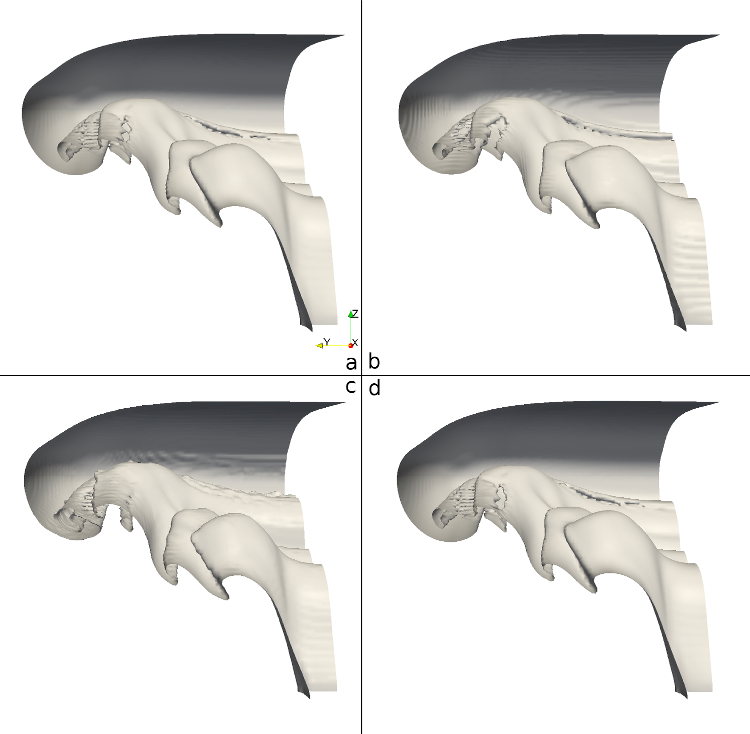} 
  \caption{The phase separation simulation, for $t\approx 1.3s.$ Pictures for methods \textbf{M1} (a), \textbf{M2} (b), \textbf{M3} (c) and \textbf{M4} (d). View in the $+x$ direction.}\label{mix13}
\end{figure}

The lighter fluid flows upwards due to simulated gravity along $z$ axis. In Figure \ref{mix13}, the shape of the interface is presented for all four investigated methods, obtained for $t=1.3$s It can be observed, that methods \textbf{M1} and \textbf{M4} yield virtually identical results. As a matter of fact, in Fig. \ref{mix13} two ``simplified'' approaches also produce interface shapes which are hardly distinguishable from the fully-fledged VOF approaches. Small disturbances in form of horizontal stripes are visible on a vertical ``oil'' core along the $z$ axis. However, the physics of the flow is reproduced consistently. It has to be noted, that in this test-case, inertia of the ``oil'' fluid is quite significant, effectively limiting the influence of any ``parasitic'' flows (produced ,as shown above, by \textbf{M3} or, to less extent, the \textbf{M2} method).

\begin{figure}[ht!]
  \centering
  \includegraphics[scale=1]{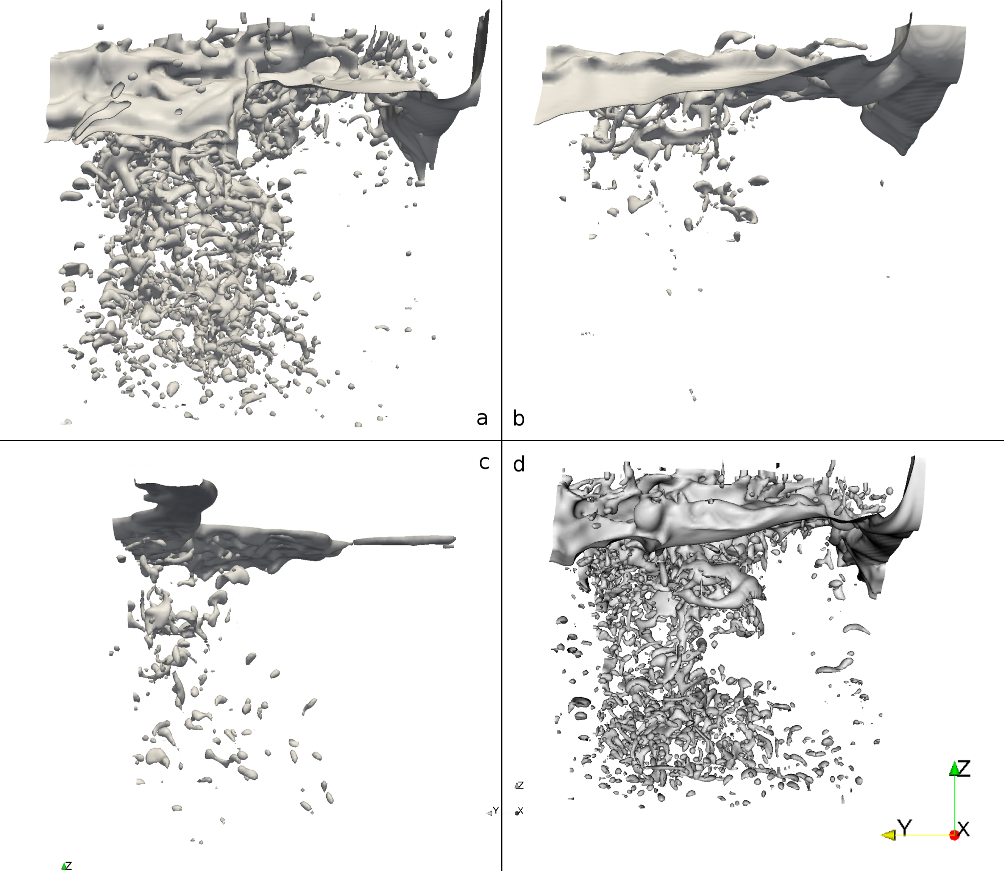} 
  \caption{The phase separation simulation, for $t\approx 7.5s.$ Pictures for methods \textbf{M1} (a), \textbf{M2} (b), \textbf{M3} (c) and \textbf{M4} (d).}\label{mix75}
\end{figure}

The turbulent nature of the flow is displayed fully in Figure \ref{mix75}, which depicts the macroscopic interface shape for $t\approx 7.5s.$ The bulk mass has undergone an overturning motion, and subsequently broke into large number of droplets, which is visible more accurately in (a) and (d) (methods \textbf{M1} and \textbf{M4}). We should add, that only the interfacial surface is visible in both Figures \ref{mix13} and \ref{mix75}, so that for example part of the domain in Fig. \ref{mix75}b is occupied by ``oil''. Taking this into account, we are still able to observe in Figure \ref{mix75}c, that \textbf{M3} has not conserved as much mass (volume) as \textbf{M2}.

\begin{figure}[ht!]
  \centering
  \includegraphics[scale=1]{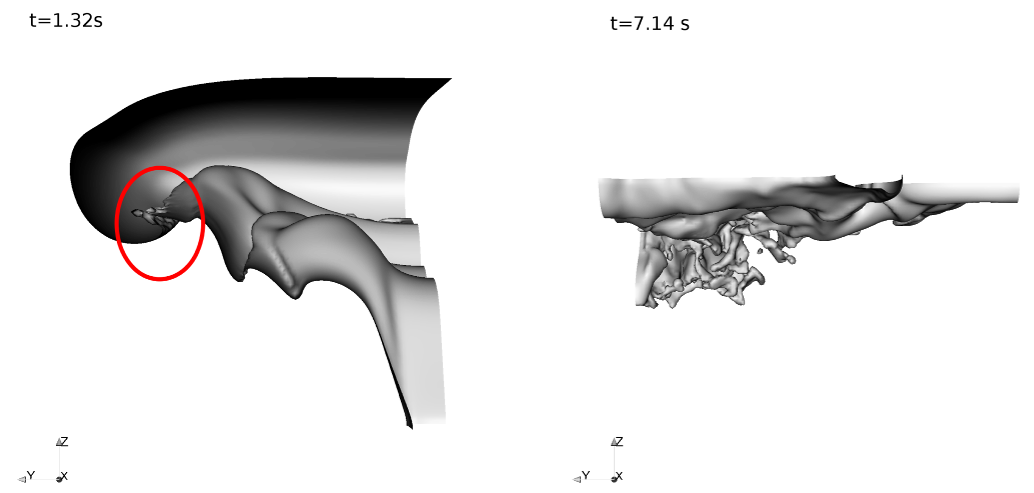}
  \caption{Macroscopic interface shapes obtained for phase separation test case using $128^3$ uniform grid, Level Set method. Images for $t=1.32$ and $t=7.14s.$ }\label{mixLS}
\end{figure}

As far as the mass conservation itself is considered, we have included results of the simulation which utilizes ``pure'' (sans VOF coupling) Level Set method in Figure \ref{mixLS}.  In the Figure, we observe that for $t=1.3$s the overall character of the interface is similar to Fig. \ref{mix13} - although one of interface ``folds'' is less resolved (encircled in Fig. \ref{mixLS}).  The right-hand side image in Figure \ref{mixLS} has been prepared for $t=7.1 s,$ in spite of which all of the droplets below the bulk mass have been lost by the LS method advection. Notably more mass is conserved even by methods \textbf{M2} and \textbf{M3} (see Fig. \ref{mix75}).

\begin{figure}[ht!]
  \centering
  \includegraphics[scale=.85]{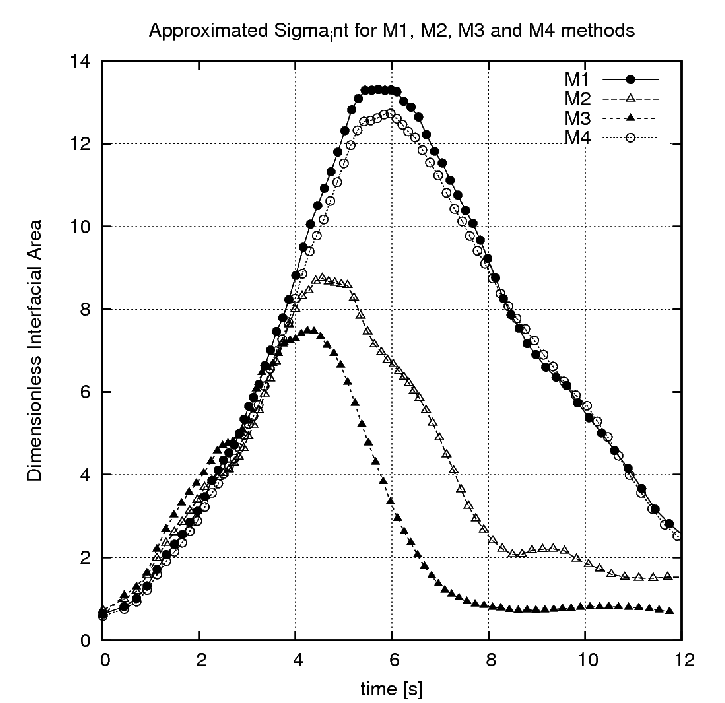}
  \caption{Evolution in time of the approximated, dimensionless $\Sigma_{int}$ interfacial area for four investigated methods.}\label{mix_surf}
\end{figure}

This visual assessment is reinforced with quantitative analysis in Figure \ref{mix_surf}, in which interfacial surface is plotted for four considered methods. The surface is approximated crudely with the expression

\begin{equation}\label{mixs}
  \Sigma_{int}(t)=\sum_{\Gamma}C_{ijk}*\Delta x * \Delta y 
\end{equation}

where $\Gamma$ signifies the interface, which, while being inaccurate in a detailed study of singular droplets, yields a reliable assessment of the large-scale tendency in the interfacial area \cite{vincent}. The curves are normalized by $\Sigma_{int}(0)$; thus we can see methods \textbf{M1} and \textbf{M4} yielding nearly 13 times the initial surface area for $t\approx 6$s, which is close to the situation depicted in Fig. \ref{mix75}. As expected from previous considerations, methods \textbf{M2} and \textbf{M3} perform much less efficient, yielding $8.7$ and $7$ times the initial area respectively. Also, the end of overturning phase of the ``oil'' is predicted $1-2$ seconds  earlier by using this methods, while the \textbf{M1} and \textbf{M4} results are in good agreements with previously published \cite{vincent, aniszewskiJCP}.

\begin{figure}[ht!]
  \centering
  \includegraphics[scale=.85]{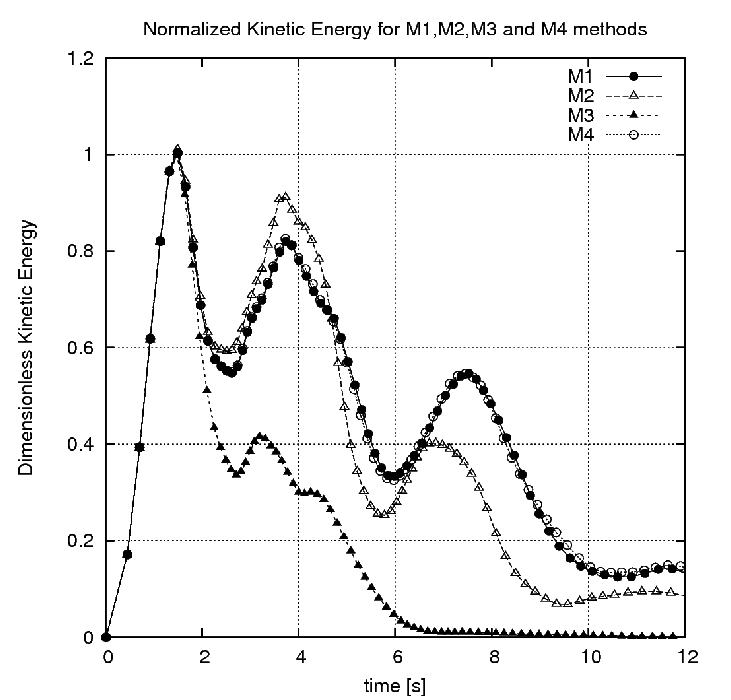}
  \caption{Evolution in time of the approximated, dimensionless kinetic energy $E_k$ in lighter fluid (oil) for four investigated methods. Normalized by maximum of $E_k$ in oil.}\label{mix_ke}
\end{figure}

We are following the analysis of $\Sigma_{int}$ by a brief examination of ``oil'' kinetic energy $E_k,$  which is presented in Figure \ref{mix_ke}. Again, results from \textbf{M1} and \textbf{M4} are not only nearly identical to each other, but also in agreement with publications \cite{vincent}. Since we have concluded before, that \textbf{M3} causes loss of traced mass, it is consistent with the fact that it also fails to predict the second peak of $E_k$ which takes place for $t\approx 4$s. This is due to the fact that most of the traced fluid has been lost, so its mass is no more taken into account. The subsequent peaks, as well as the oscillatory (``sloshing'') movements that take place later in the simulation, are not visible at all when using \textbf{M3}.

The THINC-SW method (\textbf{M2}) has performed better in this simulation; yielding correct peak placement in time for all depicted peaks; however, the second peak is significantly over- and the third under-predicted.

\subsection{Tree-dimensional Rayleigh-Tailor instability}

\begin{figure}[ht!]
  \centering
  \includegraphics[scale=2.]{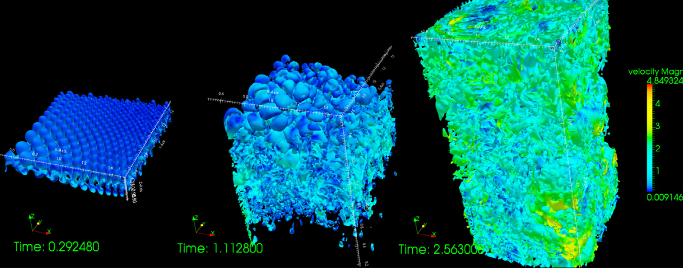}
  \caption{Various stages of Rayleigh-Taylor instability for multi-modal initial disturbance simulated using \textbf{M1} method.}\label{rtf1}
\end{figure}

Rayleigh-Taylor instability \cite{youngs1991, banerjee, chandra} occurs when in a closed space, two static fluids are superimposed with the heavier fluid on top of the lighter. In the domain with impermeable boundary conditions in all directions, two non-mixing fluids are placed, the lighter (fluid 1, F1) under the heavier F2. Thus, a pressure gradient develops in the vertical direction, with $p$ higher in the lighter fluid. Being unstable, this system is put in motion by any disturbance to the interface\footnote{In practice, numerical errors will cause that -- even if the interface is defined as flat -- albeit in an uncontrolled manner.}.  Initial position of the interface is $h(t_0)=1m.$ We impose a single mode disturbance of amplitude $h_0$ and wavelength $\alpha=1 \lb m \rb,$ so that the initial $\phi$ distribution is defined by

\begin{equation}\label{rt1}
  \phi(x)=\frac{L_z}{2}+h_0\left(\cos(kx-x_0)+\cos(ky-y_0) \right),
\end{equation}

while we choose $x_0$ and $y_0$ set to $1,$ so the disturbance is centered in the domain, when looking parallel to $z$ axis. For this and other similar kinds of initial disturbance, the flow will develop int multitude of of ``bubbles'' and ``spikes'' the first being volumes of F1 travelling upwards, and the second -- ligaments of F2 flowing downwards \cite{youngs1991}. We trace bubbles' position ($h_b$) together with spikes ($h_s$), which allows to express mixing zone thickness as $(h_b-h_s)$. The growth of mixing zone's size is covered in Taylor's original analysis \cite{taylor1950, banerjee, youngs1991}. Here we focus on the position of the bubbles. Initially, for $h_b<<1/k,$ we may expect\cite{banerjee}

\begin{equation}\label{rt2}
  h_b(t)=h_0\cosh(\Gamma t),
\end{equation}
using $\Gamma=\sqrt{A_t gk}$  where \[A_t=\frac{\rho_1-\rho_2}{\rho_1+\rho_2}\] is the Atwood number (here $A_t=\frac{1}{2}$) and $\lambda=2\pi/k.$ As the flow develops ($h_b\approx 1/k$) it transits into a regime in which a quadratic formula may be used, such as

\begin{equation}\label{rt3}
  h_b=\alpha_b A_t g t^2,
\end{equation}
where $\alpha_b$ is a parameter chosen for ''bubbles'' \cite{youngs1991}. If many bubbles exist at this stage (such as when multimode initial disturbance has been imposed) they will exhibit self-affinity, in that their growth resembles scaling (``inflating''). Regardless of the initial disturbance type, once the bubbles and spikes begin merging, and/or reach the ceiling of floor of the simulated domain, a turbulent mixing phase (see Fig. \ref{rtf1}) begins, whose character is strongly dependant of initial disturbance and Atwood number \cite{banerjee}.

For our tests, we have chosen an uni-modal initial disturbance (setting $k=2$ in (\ref{rt1})) with the aim of obtaining a less turbulent flow (compared to Fig. \ref{rtf1}), dominated by a singular, large rising ``bubble'' structure. Thus, we hoped to lessen the importance of phase fragmentation, making the simulation feasible with all four tested methods. The description of results will be divided into three subsections:
\begin{itemize}
\item Assessment of the growth rate using theoretical predictions (section \ref{rts1});
\item Macroscopic character of the flow (section \ref{rts2});
\item Remarks about mass conservation (section \ref{rts3}).
\end{itemize}

\subsubsection{Growth Rate Assessment}\label{rts1}

The growth rate assessment of the mixing zone's top (the ``bubbles'') and bottom (the ``spikes'') positions has been performed using a condition imposed on the level set function $\phi.$ Namely,

\begin{equation}\label{hb}
  h_b=max\{ z\colon |\phi(z)|<\epsilon\}
\end{equation}

and

\begin{equation}\label{hs}
  h_s=min\{ z\colon |\phi(z)|<\epsilon\},
\end{equation}

where $\epsilon$ is a small number. In effect, maximum and minimum $z$ coordinates of the interface were considered. While much less sophisticated than criteria used to measure growth area in the experimental works \cite{andrews1990}, this allows to assess whether the growth rate of bubbles using (\ref{hb}) matches the quadratic formula (\ref{rt3}).

\begin{figure}[ht!]
  \centering
  \includegraphics[scale=1.5]{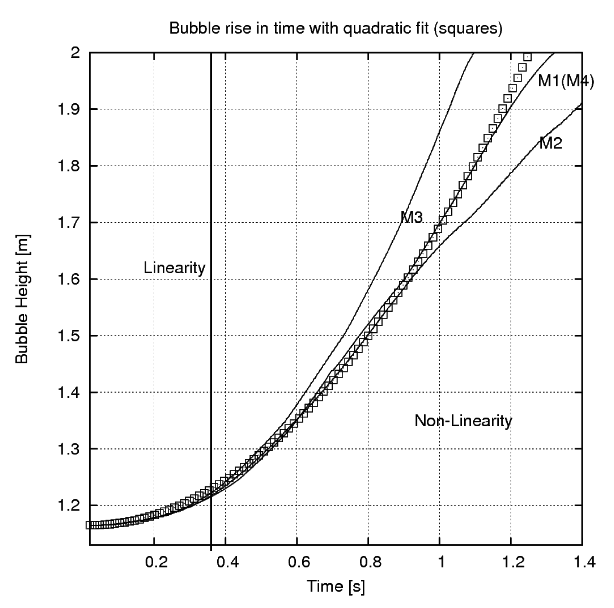} 
  \caption{Temporal evolution of the ''bubble'' height in Rayleigh-Taylor instability.}\label{rtf2}
\end{figure}

Such an assessment can be made by inspecting Figure \ref{rtf2}, which contains plots of $h_b$ for three of the tested methods. For clarity, the plot for \textbf{M4} was left out, since it was identical to \textbf{M1}. As we can see in Fig. \ref{rtf2}, the \textbf{M1}/\textbf{M4} methods produce overall good fit to a quadratic curve produced using (\ref{rt3}) (white squares). There is a slight over-prediction for $t\approx 0.8$s, for which, by contrast, the \textbf{M2} methods fits the theoretical curve perfectly. Unfortunately, for $t>0.9$s, the \textbf{M2} predicts bubble position growth much slower than both the theory and \textbf{M1}/\textbf{M4} methods.

The \textbf{M3} method generally over-predicts the $h_b,$ which is a purely numerical error. This is because in the e case of our particular simulation, it is clear that the higher $h_b$ values are due to the existence of thin interfacial formation in the centre of domain (see Figure \ref{macro_105}c) which does not appear when using other methods, nor it is visible when using less refined grids (effectively changing the \textbf{M3} curve). The bulk interface, however, undergoes evolution akin to \textbf{M2} (Fig. \ref{macro_105}b).

\begin{figure}[ht!]
  \centering
  \includegraphics[scale=1.]{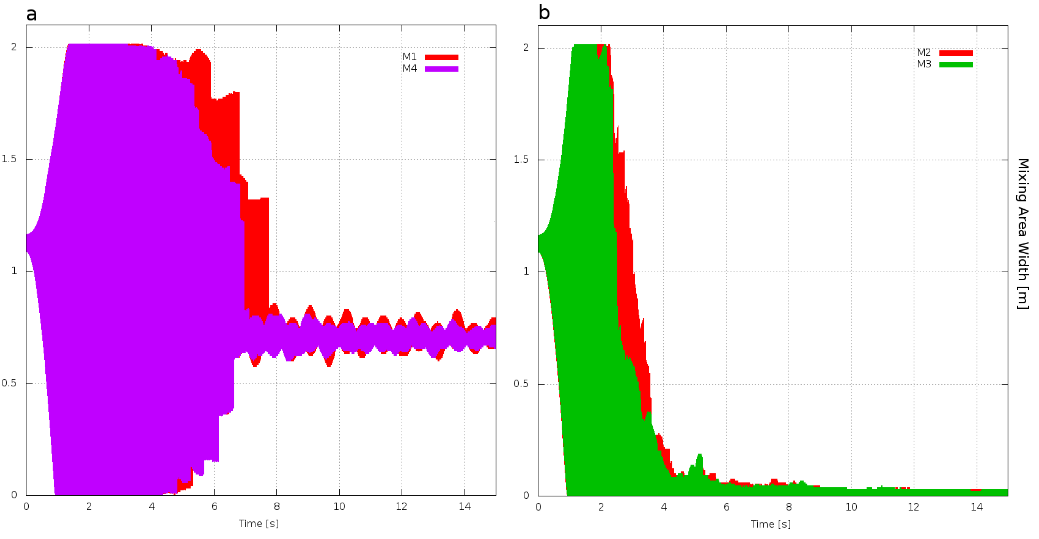} 
  \caption{The $h_b-h_s$ values plotted for methods in pairs: \textbf{M1} and \textbf{M4} (a), \textbf{M2} and \textbf{M3} (b).}\label{rtf3}
\end{figure}

In the initial, linear regime ($t<1.6$s) all the methods' predictions are identical and agree with (\ref{rt2}). Hence we do not present a separate plot for that regime.  The ``spike'' positions are also identical for all four methods - note, that $h_s>0$ only for $t<1$s, which can be seen in Figure (\ref{rtf3}), which presents $h_b-h_s$ for pairs of tested methods. Also, it is visible that only methods \textbf{M1} and \textbf{M4} predict the ''spikes'' leaving the $z=0$ level, as described in the following Section. It is tempting to interpret Figure \ref{rtf3} also as a measure of mass conservation, since intuitively, the mixing zone size should be proportional to the number of predicted ``bubble'' and ``spike'' formations. However, since formulae (\ref{hb}) and (\ref{hs}) were used to prepare Figure \ref{rtf3}, drawing mass conservation related conclusions in this basis alone is not valid, since it carries no information about the amount of mass contained the $z\in\lb h_s,h_b\rb$ volume.

\subsubsection{Macroscopic Character of the Flow}\label{rts2}

All results presented in this subsection have been prepared using a $128^2\cdot 256$ grid, calculated using 128 processor cluster on an uniform grid, amounting to a \[h=\Delta x=\Delta y=\Delta z=\frac{1}{128}\approx 7.8\cdot10^{-3}m\] spatial grid cell size.

\begin{figure}[ht!]
  \centering
  \includegraphics[scale=1.0]{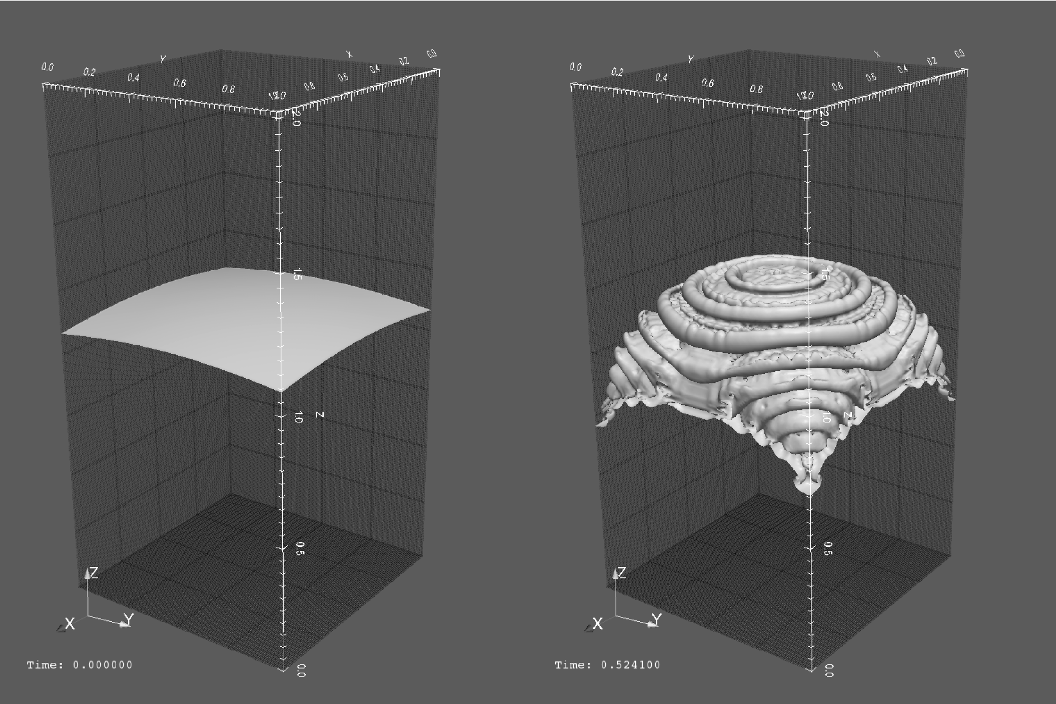} 
  \caption{Initial condition (left) and the interface shape for $t\approx 0.5$s using \textbf{M1} method (right) for the Rayleigh-Taylor simulation.}\label{macro_initial}
\end{figure}

\begin{figure}[ht!]
  \centering
  \includegraphics[scale=1.]{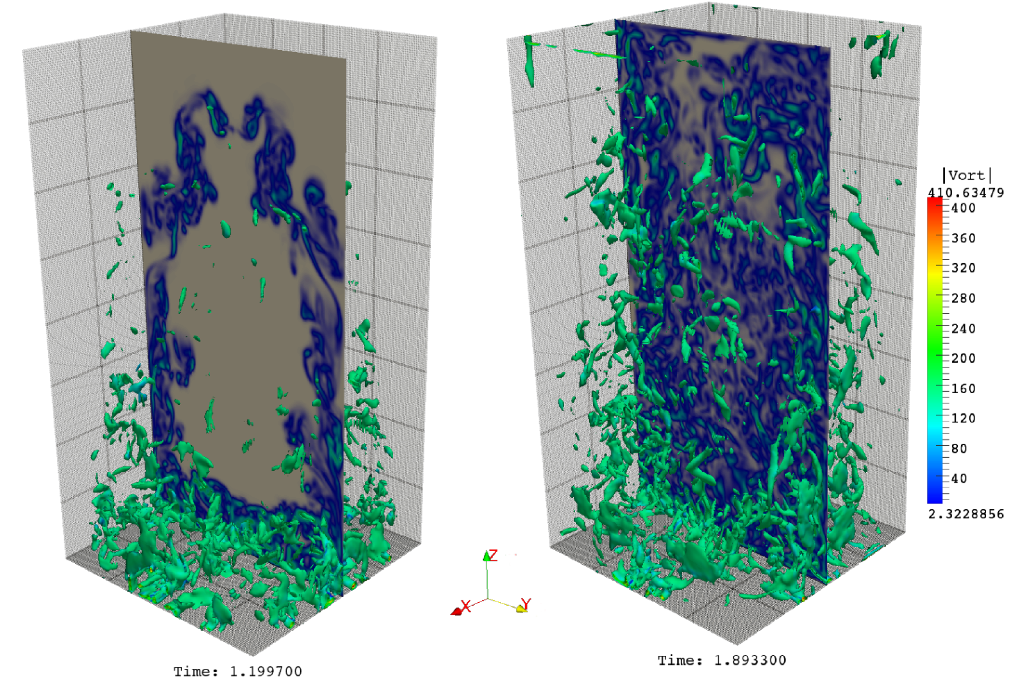} 
  \caption{Simulation of Rayleigh-Taylor instability using the \textbf{M1} method. Isosurfaces of vorticity norm and a cut-plane colored by the same quantity. Plots for $t\approx 1.19$ (left) and $t\approx 1.89$ (right).}\label{rt_vort}
\end{figure}

\begin{figure}[ht!]
  \centering
  \includegraphics[scale=1.]{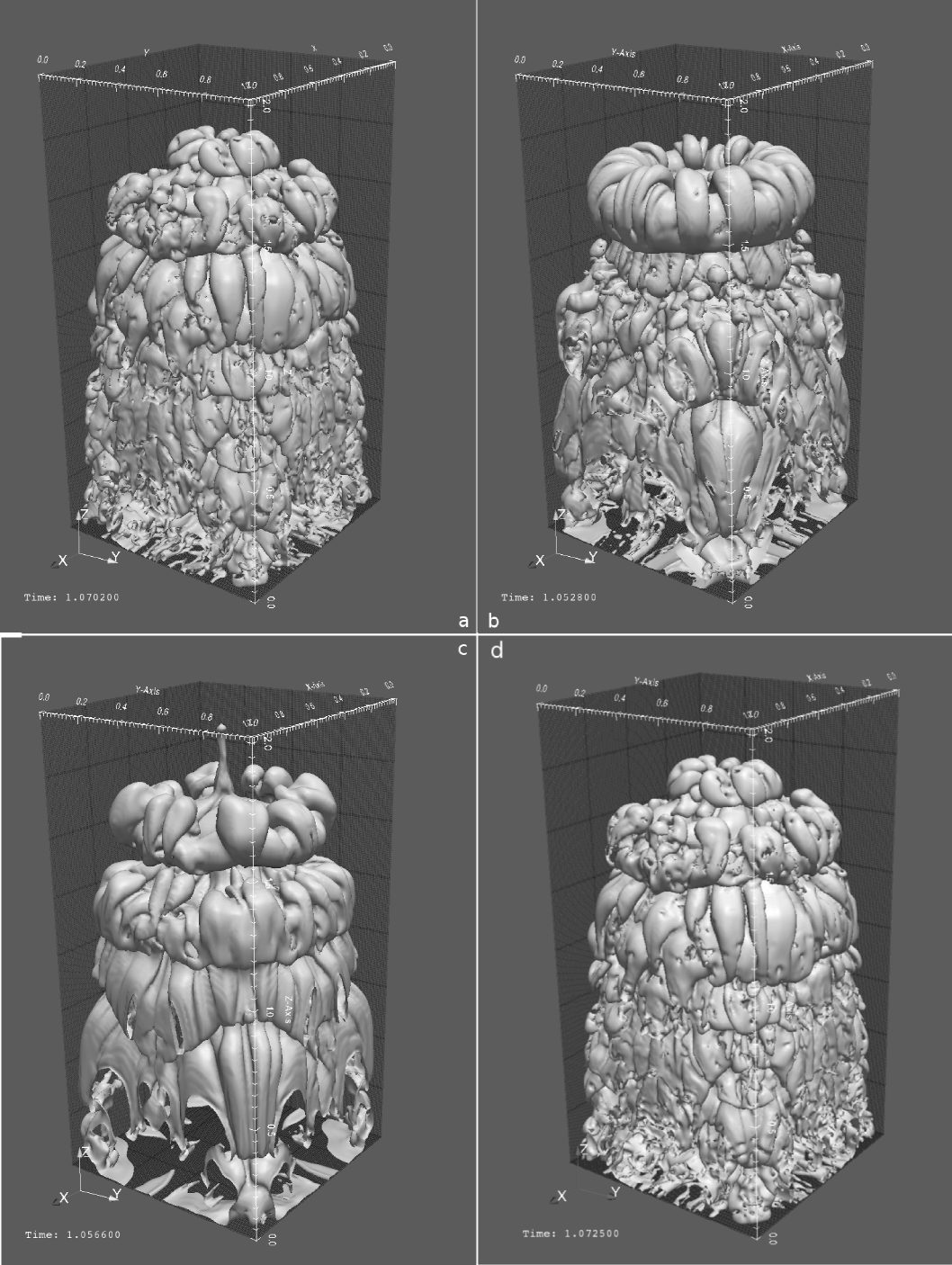} 
  \caption{Macroscopic interface shape for  $t\approx 1.05$s, using \textbf{M1} (a), \textbf{M2} (b), \textbf{M3} (c) and \textbf{M4} (d) methods.}\label{macro_105}
\end{figure}

The initial condition for the test case may be seen in Figure \ref{macro_initial}, together with an image of the interface shape obtained using \textbf{M1} for $t\approx 0.5$s. At this stage, ring-like formations appear on the surface of the rising F1 fluid ``bubble'', which may be induced numerically, e.g. by the implementation of initial condition. The fluid rise creates a multi-level structure, visible in Figure \ref{macro_105}. At this stage, the ``bubble'' formations move slower than ``spikes'' - being responsible for most of the flow inertia - that have already touched the domain bottom surface. Figure \ref{rt_vort} can be used to assess the vorticity distribution in the domain at this stage: it is clear from inspection of Fig. \ref{rt_vort}, that more vortical structures exists in the bottom half of the domain.

Even at this stage of the flow evolution we observe distinctive characteristics of methods \textbf{M2} and \textbf{M3}, such as the creation of a ''crown'' structure by \textbf{M2} (Fig. \ref{macro_105}b) and loss of ''spike'' structures by the \textbf{M3} (Fig. \ref{macro_105}c). On the other hand, \textbf{M1} and \textbf{M4} hardly differ when results are viewed at this scale (in fact, there are discrepancies at the $h$ scale).

\begin{figure}[ht!]
  \centering
  \includegraphics[scale=1.]{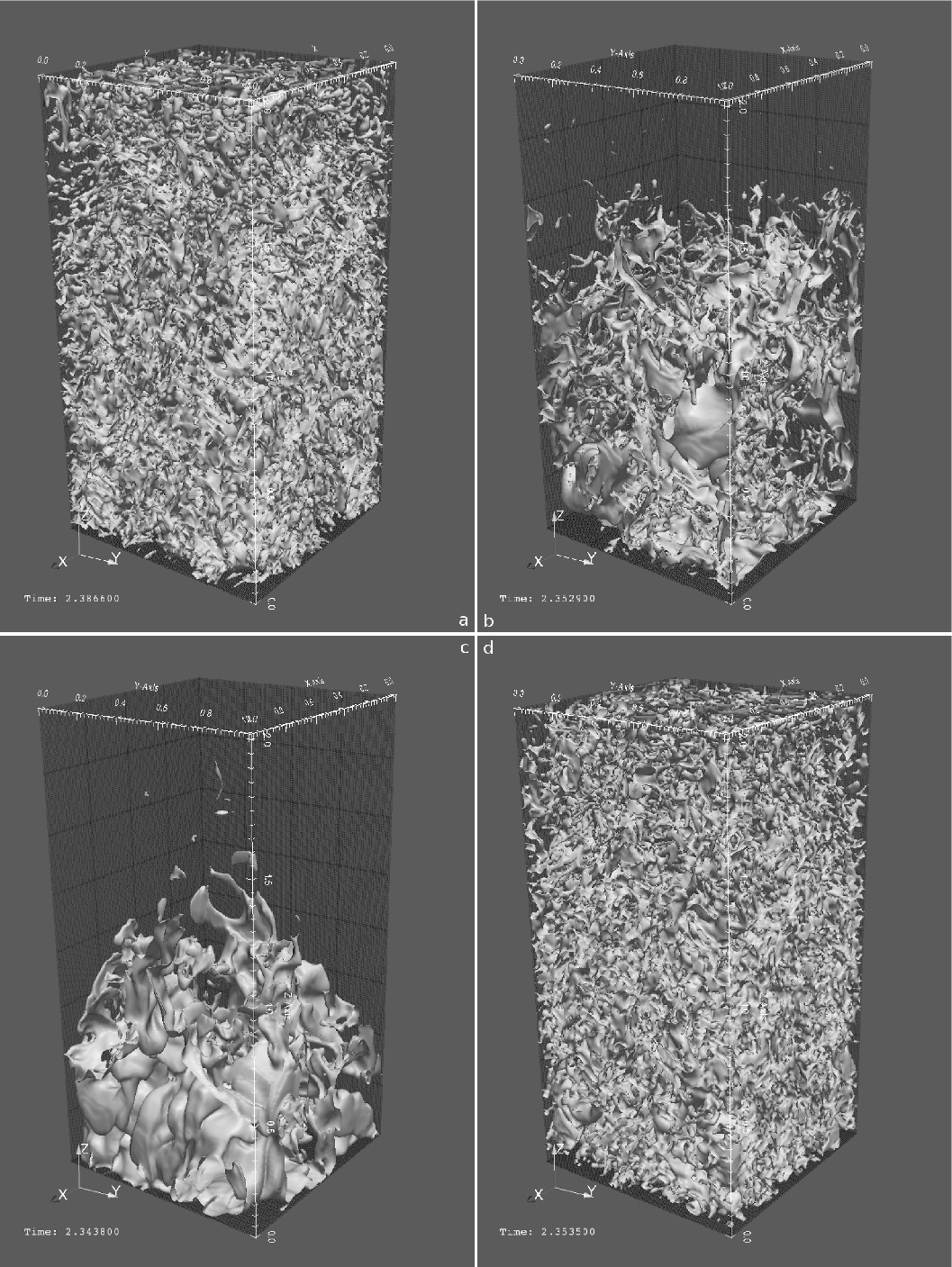} 
  \caption{Macroscopic interface shape for  $t\approx 2.35$s, using \textbf{M1} (a), \textbf{M2} (b), \textbf{M3} (c) and \textbf{M4} (d) methods.}\label{macro_235}
\end{figure}

Subsequently, the flow develops into a fully turbulent mixing, which leads to high interface fragmentation (Figure \ref{macro_235}. The \textbf{M1} and \textbf{M4} methods continue to track small interface formations, although it becomes easier to spot differences between them at this stage. The ``simplified'' methods on the other hand have by now ``lost'' part of the small bubble/spike formations, and the F1 fluid begins to fill upper part of the domain - we easily see by inspecting Fig. \ref{macro_235}c that \textbf{M3} is predicting a non-physical fluid volume proportions, since both fluids should occupy half of the domain volume. Since, in simulation of two-phase flow, both interface contributions and the spatial redistribution of inertia depend on the effectiveness of the interface representation \cite{vincent, aniszewski}, we conclude that \textbf{M2} and \textbf{M3} methods fail to predict significant features of the flow physics due to their inability to follow small structures.

\begin{figure}[ht!]
  \centering
  \includegraphics[scale=1.]{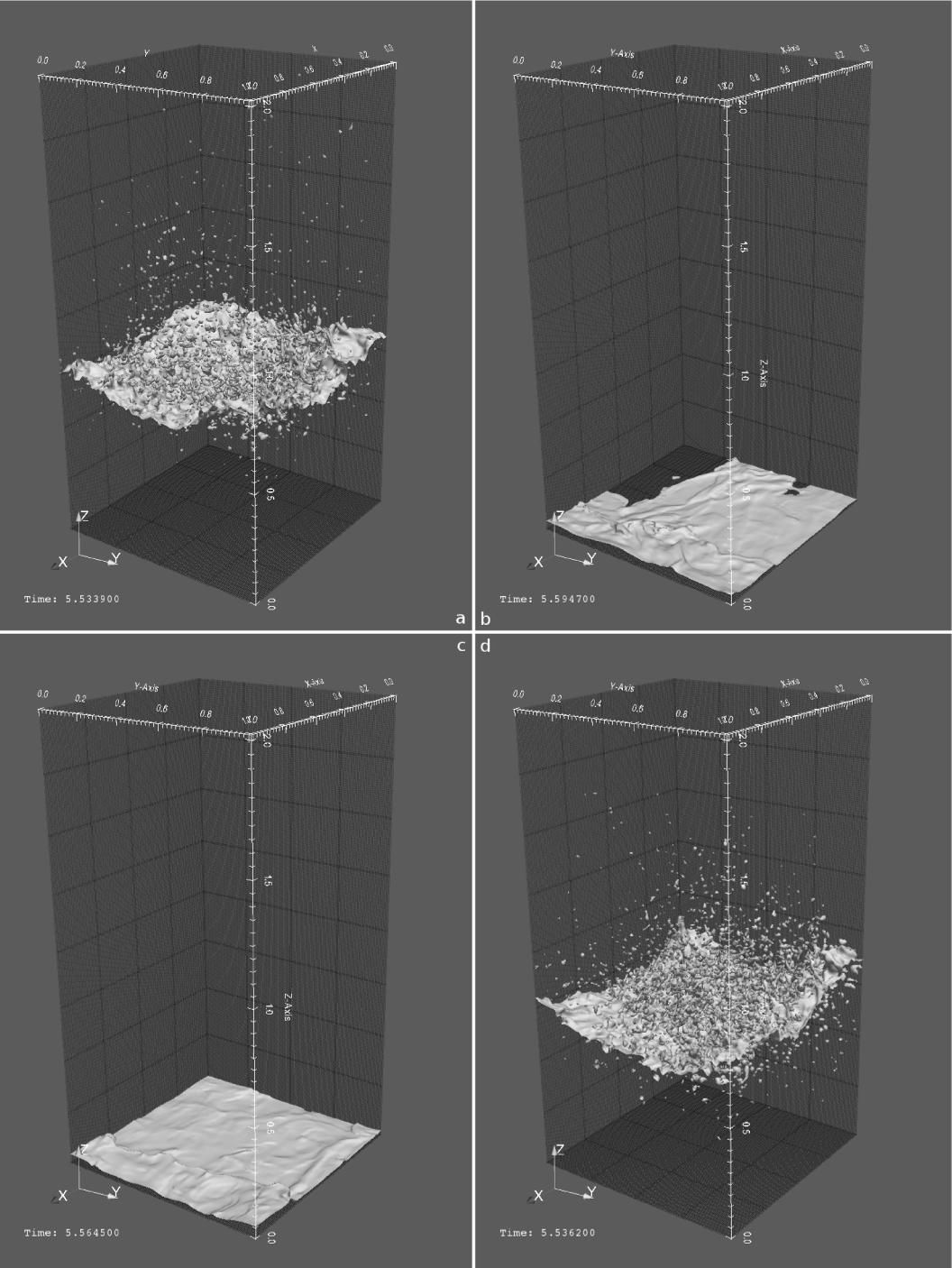} 
  \caption{Macroscopic interface shape for  $t\approx 5.53$s, using \textbf{M1} (a), \textbf{M2} (b), \textbf{M3} (c) and \textbf{M4} (d) methods.}\label{macro_553}
\end{figure}

Once the fluid mixing has largely ceased, and the bulk fluids exchanged places in the computational domain, lighter F1 should fill its upper half, placing the interface at $z=1m$. We can see in Figure \ref{macro_553} that this is not the case for neither of the methods. Of course, at $t\approx 5.5$s the interface is yet not less than $20$ seconds from becoming stationary; however, we observe that bulk interface rests at approximately $z=0.8$ meters for the \textbf{M1} and \textbf{M4} methods, meaning that nearly $20\%$ of the traced fluid volume has been lost due to numerical error. By contrast, we see that methods $\textbf{M2}$ and $\textbf{M3}$ have both led to nearly total mass loss, placing the interface at $z\approx 0.5$m, which is true for both methods even if \textbf{M3} outperformed \textbf{M2} in this aspect for the most of simulated time.

\subsubsection{Remarks Concerning Mass Conservation}\label{rts3}

\begin{figure}[ht!]
  \centering
  \includegraphics[scale=1.1]{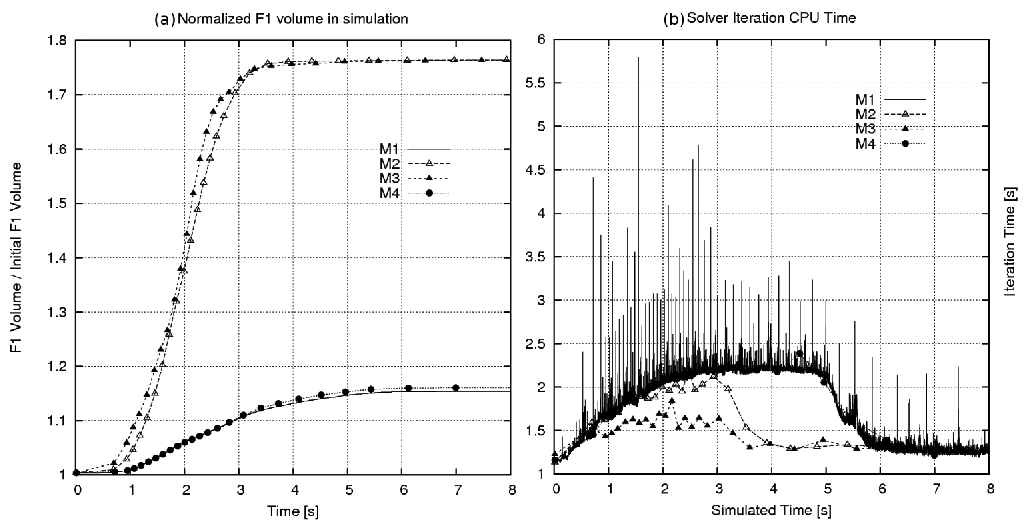}
  \caption{(a)Temporal evolution of normalized mass conservation for all four tested methods in the simulation of Rayleigh-Taylor instability using $128^2\cdot 256$ grid.(b) Solver iteration time in seconds for the same simulation.}\label{rtconsf}
\end{figure}

In Figure \ref{rtconsf}a we see plots for \textbf{M1}-\textbf{4} methods of normalized F1 volume in time. Ideal numerical methods would yield in this situation a constant function at $f(t)=1.$ However, all the methods create artificial F1 volume (that is to say ''loose'' the traced F2 volume). Curves for \textbf{M1} and \textbf{M4} are again similar, it can be seen that \textbf{M1} provides for slightly better conservation of F1/F2 balance; this is expected concerning method's high accuracy. Both methods reach level of $1.15$ meaning that $15\%$ of artificial F1 volume was created, or that $15\%$ of F2 formations were lost. We notice, that the process of interface information loss begins at $t\approx 1$s, which corresponds to the onset of mass fragmentation - at that time droplets are formed in the flow with the size comparable to $\Delta x=\Delta y=\Delta z.$ Until $t\approx 7$s, mass remains fragmented, as we also see in Fig. \ref{macro_553} (a) and (d), where small packets of both fluids are present on both sides of interface. After $t>7$s, there is no more mass loss, only the ``sloshing'' motions of the interface which are well resolved, all four curves in Fig. \ref{rtconsf}a become constant.

On the other hand, we see that methods \textbf{M2} and \textbf{M3} are not able to follow fine interfacial formations: for $t\in \lb 1,3.5\rb$ a rapid mass loss occurs leading to the loss of information about $76\%$ of F2. Quantitively, the loss is more rapid in the case of \textbf{M2}, however, asymptotically the same level of $1.76\times$initial F1 mass is reached. This shows that the ability of \textbf{M2} and \textbf{M3} to trace small (that is, not bigger than around 4 grid-cells)  droplets/bubbles within turbulent flow vortical structures and shear, is much inferior to \textbf{M1} and \textbf{M4}. 

\begin{figure}[ht!]
  \centering
  \includegraphics[scale=1.]{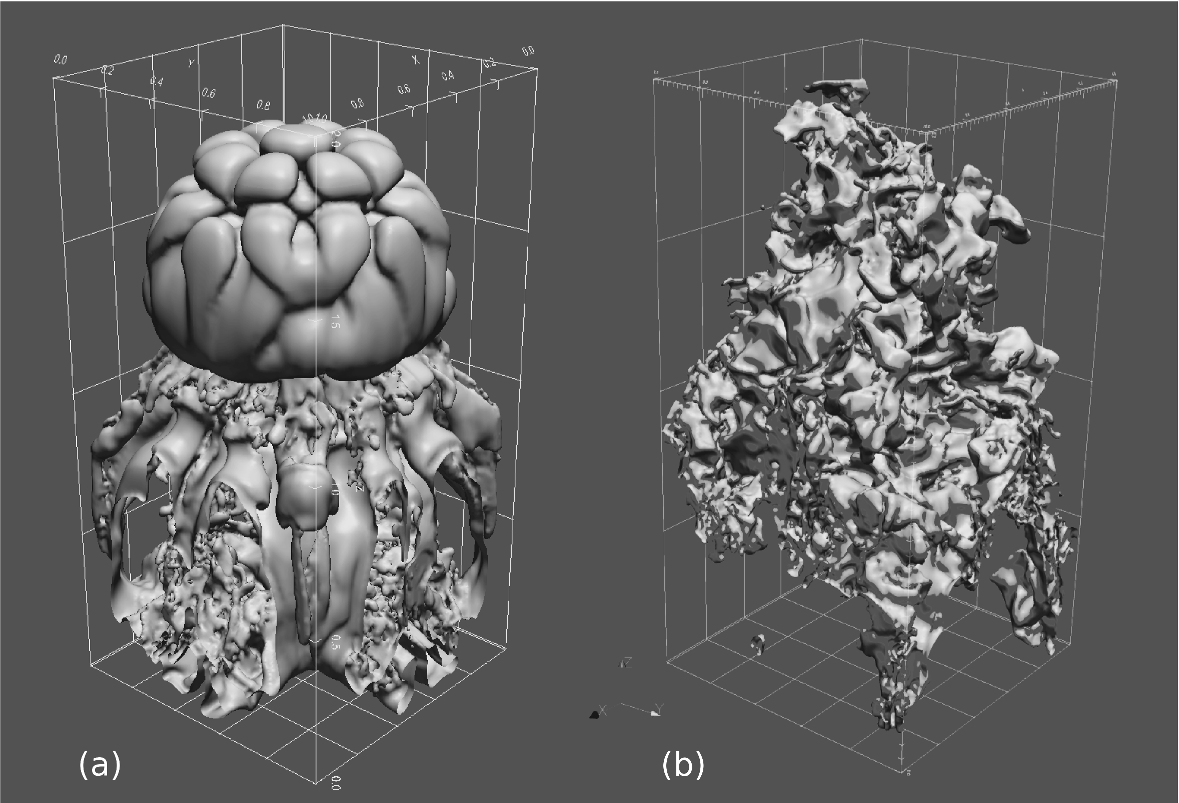}
  \caption{Macroscopic interface evolution using LS method: $t\approx 1.05s$ (a) and $t\approx 2.35s$ (b), corresponding to Figures \ref{macro_105} and \ref{macro_235}.}\label{rtlspure}
\end{figure}

We conclude the discussion concerning mass conservation by focusing the attention on Figure \ref{rtlspure}, which presents results of simulating the Rayleigh-Taylor instability using ''pure'' Level Set method. Since using LS method we obtain generally smoother surfaces \cite{osher2000} than using CLSVOF, initial interface shape contains less wrinkles (than what could be observed in Fig. \ref{macro_initial}). As a consequence, the bulk fluid in Fig. \ref{rtlspure}(a)  does not contain distinguishable ``crown'' formations, although its position resembles that in Fig. \ref{macro_105}.  However, less small-scale features is obtained at $t\approx 1.05s$ using Level Set method visible in Fig. \ref{rtlspure}(a), than e.g. the \textbf{M2} method in Fig. \ref{macro_105}(b) near the bottom of the domain ($0<z<1$). For $t\approx 2.35s$ in Figure \ref{rtlspure}(b), we observe less fragmentation than using \textbf{M2} method (Fig. \ref{macro_235}(b), with the result being comparable to \textbf{M3} (Fig. \ref{macro_235}(c)), albeit the interface is more shifted in $z+$ direction. Less individual droplets are visible in Fig. \ref{rtlspure}(b) than using \textbf{M2}, although comparison with \textbf{M3} would again be nontrivial. It is clear that the \textbf{M3} method yields a result closest to LS in this particular test; however, results such as Fig. \ref{mix75} and \ref{mixLS} presented before keep us convinced that this method may still improve mass conservation.  Again, inspecting Fig. \ref{rtlspure} there is no real competition between Level Set simulation result and those obtained ising \textbf{M1} and \textbf{M4} (Figure \ref{macro_235}(a)(d)). 

\subsubsection{Remark Concerning CPU Time}

In the simulation of Rayleigh-Taylor instability, the \textbf{M1} method was ``slowest'' in terms of CPU time, while the ``fastest'' of them has been \textbf{M3}, judging by the averaged CPU time needed to compute a single solver step (see Figure \ref{rtconsf}b. In this Figure, plot of $t_{end}-t_{start}$ (using MPI subroutines) has been plotted for all four tested methods. The plot for \textbf{M1} has been left unchanged while other methods have only each $1000$th value plotted. Note that the plotted values are not $dt$ timesteps used in temporal discretization, but the wall-clock CPU time\footnote{Although, since the computation was performed using a 128-core machine, this should be understood as a 'distributed' time, that is, an interval between the moment when first of processes entered time step $n$ and the moment when the last process left it for $n+1$-st.}, thus, we see that solver took generally most of computational time for calculations in the $t\in\lb 0,5\rb$ interval, when the flow was turbulent. Noticeably, the \textbf{M3} method yielded shortest iteration times. Performance of the \textbf{M4} was comparable to \textbf{M1} (full dots signifying \textbf{M4} are almost covered with the plot of \textbf{M1}).

Among the ''simplified'' methods, the \textbf{M2} proved slower than \textbf{M3}. This tendence of  \textbf{M2} to require much CPU time might be atributed to the fact that - when programmed in a straightforward manner sans optimisation - it relies on FORTRAN intrinsic functions to compute hyperbolic functions in (\ref{th4}): this is freqently seen as an expensive operation by FORTRAN programmers. On the other hand, good performance \textbf{M3} (CPU-wise) may be due to relative simplicity, small number of instructions and lack of intrinsics. But this judgement may be biased by \textbf{M3}'s inclination to cause mass loss, which leads to certain flow simplification, and subsequently lowers CPU requirements.

\section{Conclusions}

Basing on the preceding tests, we conclude that all of the tested methods could be considered for use in modelling two-phase flows, depending of the degree of complication that a given flow exhibits. Methods \textbf{M1} and \textbf{M4} have been used in many applications (e.g. \cite{menard,duret,aniszewskiJCP}), therefore we do not feel entitled to give any assessments and conclusions about them. The differences between them lie mainly in the area of implementation; however we have supplied new material proving that their results can be validated, as is the case in Fig. \ref{rtf2}, or that they are equivalent to machine accuracy, as in Figure \ref{fc} or Table \ref{convtable}. This does not make them fully equivalent, as Figure \ref{oscillation1} shows.

''Simplified'' methods \textbf{M2} and \textbf{M3} are not that well established. Our tests conclude that method \textbf{M2} (the THINC-SW) is a very good balance between capabilities and the required number of code lines; the VOF flux computation part takes less than 20 lines of FORTRAN code, compared to over a thousand in Menard et al. \cite{menard}. Its mass conservation performance is impressive for such a simple approach. It is reliable and  rarely produces  non-physical effects, its errors are bounded even on coarse grids, as in the static droplet case.  Coupling of THINC with Level Set  yields better mass conservation than ``pure'' LS advection, as visible in Figures \ref{serpentine1} and \ref{mixLS} or \ref{macro_105}/\ref{rtlspure}. This constitutes a convincing argument for research teams  wishing to improve their Level Set simulations in terms of mass conservation, or researchers performing non-geometrical VOF calculations \cite{wacl_turbulence} that may suffer from poor mass conservation. The \textbf{M2} is a very simple way for a research team to quickly begin introducing eulerian VOF methodology into their codes.

Method \textbf{M3} -- the WLIC (or ``SVOF'' \cite{svof}) scheme -- has performed minimally worse than \textbf{M2} in terms of mass conservation. In Rayleigh-Taylor instability simulation - which involved the highest degree of interface fragmentation - the \textbf{M2} was just as unable to conserve mass as the \textbf{M3}. We have found erroneous behaviour when simulating droplet oscillation using \textbf{M3}. This method is also a little more complicated than \textbf{M2} (requiring around 70 lines of FORTRAN code to implement), and it requires all components of interface normal $\nb$ to compute the fluxes. At the same time, method \textbf{M2} requires only one component, $n_x$, so remains simpler to implement. With all its shortcomings, the WLIC remains a major simplification over PLIC VOF. The free term $\alpha$ in interface description is not needed, and fluxes are found in a straightforward manner. It also rests more (than THINC/SW) on geometrical considerations, being  a good way to transit the code to PLIC. Finally, works of Yokoi \cite{yokoi2013} prove that given appropriate curvature calculation routines in a non-CLSVOF arrangement, WLIC can produce very useful results.

Note that none of the tested ''simplified'' methods are on par with full CLSVOF/PLIC implementations \textbf{M1} / \textbf{M4}. It implies, that THINC/SW and WLIC methods -- if they are to be used to simulate surface tension dominated flows (in which, great care is needed not to affect the curvature calculation by introducing advection errors), or flows with high interface fragmentation (such as atomization or complex breakups) -- will require dense grids to compensate for their relatively low accuracy and tendency to loose the traced mass/reconstructed volume. As shown in Table \ref{convtable} especially the \textbf{M3} (WLIC) has convergence inferior to \textbf{M2}, and effectively provides much less resolution. Although the results of the static droplet test were not discouraging (Table \ref{statictab}), both these methods' performance in the droplet oscillation test-case \ref{droposc}  indicates, that either greater grid resolution or/and  an entirely different curvature/surface tension calculation scheme \cite{brackbill,yokoi2013} might be vital if such simulations are ever to yield acceptable results. 

We are aware, that having tested the four methods as a part of a CLSVOF setup, we are only partly able to assess their influence on the final flow characteristics. This particularly concerns the conservation of traced mass/reconstructed volume. Had the tests been conducted without coupling VOF with LS, total mass conservation might become a less pressing matter, with PLIC, THINC/SW and WLIC methods producing similar results as far as global $C$ sums were considered. However, to compare them with Level Set results, one would have to consider volumes, which would have to be reconstructed -- thereby bringing us back to the setup chosen in this work.  Similarly, the LS-specific curvature calculation errors would not influence a purely VOF-based approach; however it might be argued that in this work, all tested methods used identical curvature calculation scheme.

  Also, the setups which use only VOF (and not Level Set) to follow the interface are not easily realizable without grid refinement \cite{popinet2}, or are forced to use various flavors of CSF scheme to compute curvature \cite{yokoi2013,vincent} -- techniques which may just as well introduce their own specific errors. Taking it into account, it seems just as beneficial to compare the behavior of the tested methods as parts of CLSVOF setup, advantageous also by availability of the pure LS advection results as a reference point.

\section{Acknowledgements}
Parts of this work were supported by the ANR (L'Agence Nationale de la Recherche) project MODEMI (Mod\'{e}lisation et Simulation Multi-\'{e}chelle des Interfaces, ref. ANR-11-MONU-0011). The work of M. Marek has been performed within a statutory research project BS/PB-1-103-3010/11/P.

The authors would like to thank T. Wac\l awczyk and O. Desjardins, as well as other persons whose questions and ideas have helped to make this article better. We thank the anonymous reviewers of the article for their valuable suggestions -- such as creating subsection \ref{droposc} and other ideas that have, hopefully, broadened the scope of this paper.

We express our gratitude to the CRIHAN (Centre de Ressources Informatiques de HAute-Normandie, \texttt{www.crihan.fr}) computational centre, whose computational resources we have utilized. All the visualizations have been prepared using Paraview \cite{kenneth} and Gnuplot \cite{gnuplot}.


\bibliography{4v}
\end{document}